\documentclass[aps,prl, twocolumn, superscriptaddress, longbibliography]{revtex4-2}

\usepackage{times}
\usepackage{amsmath}
\usepackage{amssymb}
\usepackage{graphicx}
\usepackage[caption=false, position=top, singlelinecheck=off, justification=raggedright]{subfig}
\usepackage{color}
\usepackage{pst-node}
\usepackage[unicode]{hyperref}
\hypersetup{unicode=true, colorlinks=true, linkcolor=blue, citecolor=blue}

\usepackage[normalem]{ulem}

\usepackage{mhchem}

\begin{document}

\title{Quantum rotation sensor with real-time readout based on an atom-cavity system}
\author{Jim Skulte}
\affiliation{Zentrum f\"ur Optische Quantentechnologien and Institut f\"ur Laser-Physik, Universit\"at Hamburg, 22761 Hamburg, Germany}
\affiliation{The Hamburg Center for Ultrafast Imaging, Luruper Chaussee 149, 22761 Hamburg, Germany}

\author{Jayson G. Cosme}
\affiliation{National Institute of Physics, University of the Philippines, Diliman, Quezon City 1101, Philippines}

\author{Ludwig Mathey}
\affiliation{Zentrum f\"ur Optische Quantentechnologien and Institut f\"ur Laser-Physik, Universit\"at Hamburg, 22761 Hamburg, Germany}
\affiliation{The Hamburg Center for Ultrafast Imaging, Luruper Chaussee 149, 22761 Hamburg, Germany}

\date{\today}

\begin{abstract}   
Using an atom-cavity platform, we propose to combine the effective gauge phase of rotated neutral atoms and the superradiant phase transition to build a highly sensitive and fast quantum rotation sensor. The atoms in a well-controlled array of Bose-Einstein condensates are coupled to a single light mode of an optical cavity. The photon emission from the cavity indicates changes in the  rotation frequency in real time, which is crucial for inertial navigation. We derive an analytical expression for the phase boundaries and use a semi-classical method to map out the phase diagram numerically, which provides the dependence of the photon emission on the rotation. We further suggest to operate the sensor with a bias rotation, and to enlarge the enclosed area, to enhance the sensitivity of the sensor.
\end{abstract}

\maketitle

Quantum sensing aims to take advantage of the accuracy and long-time stability of quantum platforms, such as ultracold atoms, for new technological applications \cite{Amico2021}. While optical Sagnac interferometers are important tools in present day navigation, they can fall short due to limited sensitivity and long-term stability \cite{lefevre1994}. Atom interferometry emerged as a promising platform to address these issues with high precision rotation and acceleration measurements \cite{Durfee2006,Burke2009,Hamilton2015,Bongs2019,Moan2020,Woffinden2022}. The applications range from testing fundamental physics \cite{Bouchendira2011,Fixler2007,Parker2018}, metrology \cite{Rosi2014}, absolute gravimetry \cite{Freier2016,Karcher2018} to inertial navigation \cite{Grewal2020,Geiger2020}.
So far, the usage of atom interferometers as rotation sensors to measure time-varying signals has been challenging. Each measurement is done destructively, which means that any change in the rotation that occurs during the preparation of the new measurement is not detected.  An interleaved operation can be used to increase the repetition rate \cite{Savoie2018}. However, an accurate rotational measurement of a time-varying rotation with an atom-only interferometry remains challenging, but is crucial if the device is to be used for inertial navigation \cite{JEKELI2005}.  \\
\indent Rotating an ultracold atom system creates an artificial gauge field \cite{Bhat2006,Rey2007,Nunnenkamp2008}, which was proposed to study gauge field-driven atom dynamics in a controlled environment and to produce cat states for quantum metrology \cite{Hallwood2006,Rey2007}. A similar setup described by a rotating Bose-Hubbard model has been proposed for rotation sensing \cite{Jiang2022}. \\
\begin{figure}[!htpb]
\centering
\includegraphics[width=1\columnwidth]{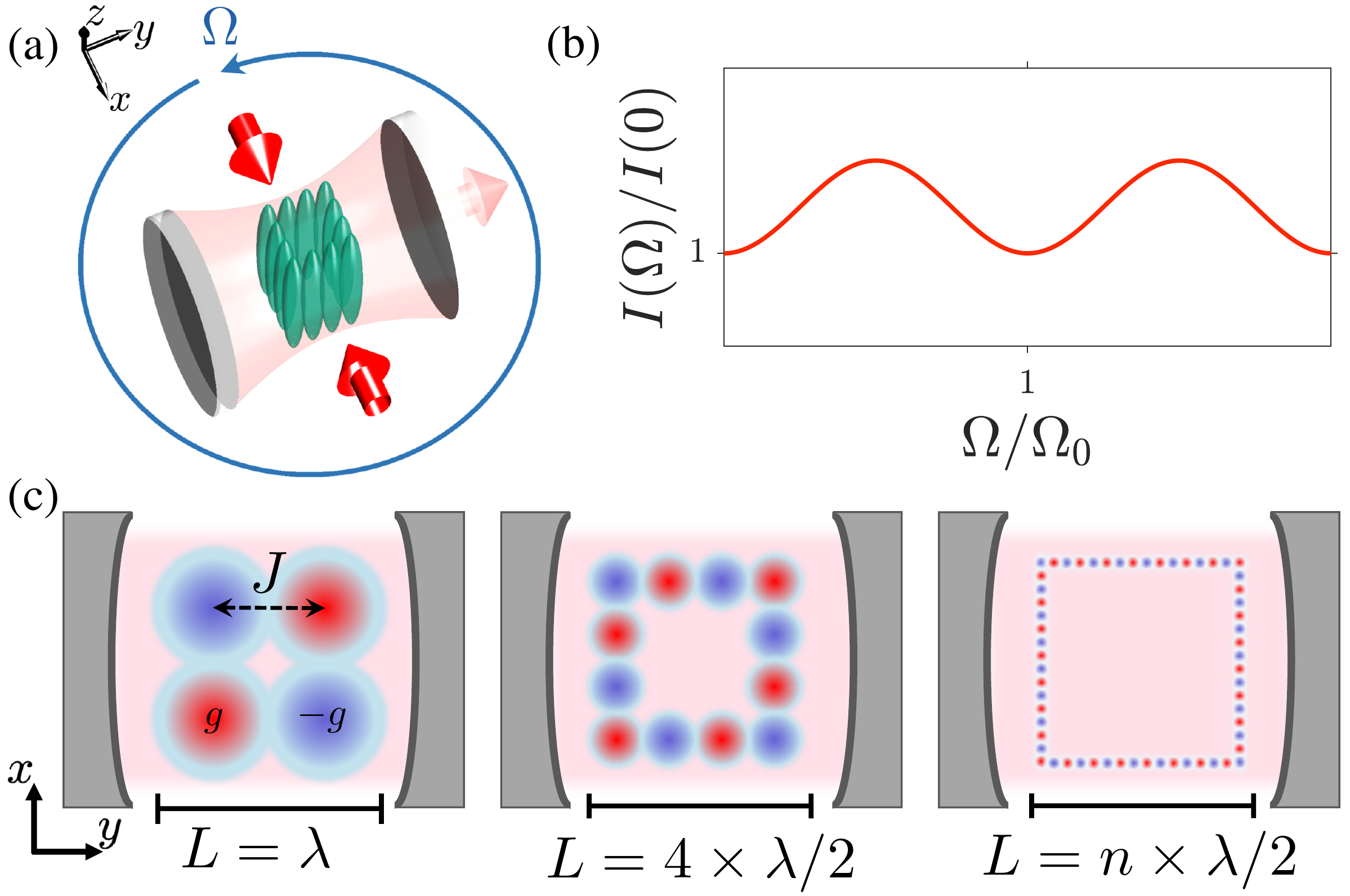}
\caption{(a) Sketch of the proposed system. 1D tubes of Bose-Einstein condensates are placed with a separation of $\lambda/2$ in a high-finesse optical cavity. The atoms are transversely pumped by a red-detuned laser beam with wavelength $\lambda$ that forms a standing wave potential. The effect of an applied rotation around the $z$-axis at an angular frequency of $\Omega$, as sketched in blue, can be measured in real time through the intensity of the photons transmitted out of the cavity at a rate $\kappa$, as shown in (b). (b) A schematic sketch of the measured light field intensity for varying rotational frequencies $\Omega$ rescaled by the intensity measured when the system is at rest. (c) Top view of the proposed setup. Red and blue circles denote the sign of the light-matter coupling $\pm g$ at each site. $J$ denotes the tunnelling between neighbouring sites. Using a DMD, a square annulus light potential is applied shown in light blue. From left to right: the proposed setup with increasing area enclosed by the atoms. This allows for an increased frequency sensitivity of the sensor, which scales with the number of array sites, $1/M$.} 
\label{fig:1} 
\end{figure}
\indent Open quantum systems such as atom-cavity systems \cite{Ritsch2013,Mivehvar2021} allow for a real-time monitoring of the photons emitted out of the cavity. This has been used to in-situ observe phase transitions \cite{Baumann2010,Klinder2015_2,Vaidya2018,Kongkhambut2022,Dreon2022}. The precise control of cold atom experiments combined with the inherent cooling of the system due to energy leaking out of the cavity, makes it an ideal platform to study non-equilibrium dynamics \cite{Kollath2016,Mivehvar2017,Bentsen2019,Holland2020,Skulte2021,Rodrigo2022,Zhang2022,Bernardis2021}.  \\
\indent In this letter, we propose to build a quantum rotation sensor that allows to measure fast time-varying signals by incorporating atom interferometry, artificial gauge phases in ultracold atoms, and real-time readout of open quantum systems. Specifically, we propose to combine the control of the hopping parameter $J$ by an effective gauge phase, which can be realized by rotations of neutral atoms, and the superradiant (SR) phase transition that can be realized by coupling neutral atoms to a single mode cavity with a photon loss rate $\kappa$. With this, we propose to create a quantum rotation sensor that possesses both non-destructive detection and real-time observation of precise changes in the external rotational frequency $\Omega$ of the system. \\
\indent Using an mean-field ansatz, we analytically show the dependence of the phase boundary and the intracavity photon number on the rotation, represented by a gauge phase. We further use a semi-classical method to numerically determine the phase diagram as a function of the rotation frequency, and the light-matter coupling strengths. We explore the dynamics under dynamical changes of the rotation. These changes can be distinguished dynamically from the inherent quantum noise, which we quantify as the standard deviation of the light field fluctuations. We note that the timescale on which changes of the phase cannot be dynamically distinguished is on the order of the inverse of the hopping parameter, which corresponds to $1/J  \approx 5\times 10^{-4}~\mathrm{s}$, for the parameters given below.\\
\indent A sketch of our proposed sensor is shown in Fig.~\ref{fig:1}(a) and with a top view depicted in Fig.~\ref{fig:1}(c). We propose to split a Bose-Einstein condensate of \ce{^87Rb} atoms into $M$ one-dimensional (1D) tubes and organize them in a two-dimensional square array with a spacing of $\lambda/2$, where $\lambda$ is the wavelength of the laser beam used to produce the light-matter coupling between the atoms and the cavity. To trap the atoms and produce the atomic array, a DMD can be employed to construct a square annulus potential with barriers running along each side at a periodicity of $\lambda/2$ to produce an atomic array \cite{Gauthier2016,Woffinden2022}. This system is then placed inside a high finesse cavity with a loss rate $\kappa$. In Fig.~\ref{fig:1}(c), we show schematic diagrams of the desired geometry of the atomic array including the confining potential. The enclosed area can be enlarged leading to an improved frequency sensitivity that scales with $1/L$, a typical strategy done in atom interferometers \cite{Moan2020}. This sensor allows to measure changes in the rotational frequency via the light intensity transmitted out of the cavity. The mechanism crucial for the operation is the following. For the SR phase to occur, the light-matter coupling strength needs to exceed the tunnelling energy, which is minimized if all sites have the same population (see Eq.~\ref{eq:E}). However, the SR phase relies on an imbalance between neighbouring sites as is depicted in red and blue in (c). Rotations of the system lead to an effective gauge field that modifies the tunnelling amplitude. This leads to an effective reduction of the real part of the tunnelling energy, which lowers the critical light-matter coupling needed to enter the SR phase. For a fixed light-matter strength the rotational frequency can now be read off by the light field intensity leaking out of the cavity. This effect is used in our rotational sensor. We note that the sensitivity is further increased with an operation of the sensor close to the phase transition. This allows for small changes in the rotational frequency induce large changes of the photon number, as demonstrated below, due to the strong dependence near the phase transition.\\
\indent The atomic sector of our system is described by a bosonic field operator $\hat{\Phi}(\mathbf{r})$ and the many-body Hamiltonian
\begin{equation}
\label{eq:0}
\hat{H}_\mathrm{A}= \int\hat{\Phi}^\dagger\left( \textbf{r}\right) \left(-\frac{\hbar^2 \nabla^2}{2m}-\Omega \hat{L}_z+V(\textbf{r})\right)\hat{\Phi} \left( \textbf{r}\right)   d \textbf{r},
\end{equation}
where the external $\hat{L}_z$ is the angular momentum operator and $V(\mathbf{r})$ is the external potential shaped by a DMD \cite{Gauthier2016,Woffinden2022}. A DMD can be used to generate a square array of potential wells, such that the atoms are confined with a width of $\lambda/2$ as sketched on Fig.~\ref{fig:1}(c). We further assume that the external potential will have a periodicity of $\lambda/2$ \cite{Woffinden2022} along the $x-$ and $y-$direction and is sufficiently weak such that the system remains in the superfluid regime \cite{Bloch2008}. This condition places a constraint on the length of each side $L$ to be an integer multiple of $\lambda/2$. The cavity photon annihilation (creation) operator is $\hat{a}$ ($\hat{a}^\dagger$) and the corresponding Hamiltonian is $\hat{H}_\mathrm{C} = \omega \hat{a}^\dagger \hat{a}$. The atoms and photons interact according to  \cite{Ritsch2013}
\begin{equation}
\hat{H}_{\mathrm{LM}}=\int\hat{\Phi}^\dagger \left( \textbf{r}\right) \eta \cos(kx)\cos(ky) \left( a+a^\dagger \right) \hat{\Phi} \left( \textbf{r}\right)  d \textbf{r},
\end{equation}
where $\eta$ the strength of the coupling.
We further assume that the potential $V(\mathbf{r})$ is deep enough to neglect next-nearest neighbour tunneling and that the band gap is larger than the rotational energy. With this, we expand the atomic field operators in Wannier orbitals \cite{Nunnenkamp2008}
\begin{equation}
\tilde{W}_i = \exp\left(- \frac{i m}{\hbar} \int_{\textbf{r}_i}^{\textbf{r}_{i+1}} \textbf{A}(\textbf{r}')d\textbf{r}'\right)W_i(\textbf{r}),
\end{equation}
where $\textbf{A}(\textbf{r}) = \Omega ~\textbf{z} \times \textbf{r}$ is the effective vector potential induced by the rotation. \\
In the rotating frame, the effective Hamiltonian for the system is given by
\begin{align}
\hat{H} =& \omega ~\hat{a}^\dagger a- g \left(\hat{a}^\dagger+\hat{a}  \right)\sum_{i=1}^M\left(-1\right)^i \hat{n}_i \\ &-J \exp\left(i \theta\left(\Omega\right) \right) \sum_{i=1}^M\left(\hat{b}^\dagger_i\hat{b}_{i+1} \right)+\mathrm{h.c.} \notag
\label{eq:1}
\end{align}
with the periodic boundary condition $\hat{b}_{M+1}=\hat{b}_1$. $J$ is the tunneling energy between neighboring condensates, and $g$ is the light-matter coupling for the Wannier orbitals. These are defined as $ J = \int d \textbf{r} W^*_i\left[-\frac{\hbar^2 \nabla^2}{2m}+V(\textbf{r}) \right]W_{i+1}$ and $g =\eta \int d \textbf{r} W^*_i  \cos(kx)\cos(ky) W_i$.
The number operator at site $i$ is $\hat{n}_i=\hat{b}^\dagger_i\hat{b}_i$, and $\theta$ is an effective phase generated by the gauge field $\theta  = \int_{x_i}^{x_{i+1}} \textbf{A}(\textbf{r})d\textbf{r}=\pi^2 n_\mathrm{s} \Omega/\omega_\mathrm{rec}$. For the four-site model shown in Fig.~\ref{fig:1}(c), this phase is the same four the four bonds. For larger realizations, shown in Fig.~\ref{fig:1}(c) as well, the gauge phase will be dependent on the bond. However, the general functionality is retained. The recoil frequency due to the pump is  $\omega_\mathrm{rec}=\hbar k^2/2m$ and we define the number of sites on each side as $n_\mathrm{s}=M/4+1$. We note that other geometries, e.g. rectangular potentials instead of a square potential, can be used as well. Here, the angle $\theta$ will be different for the $x-$ and $y-$axis. However, the functionality is retained for this modification as well, even for the four-site realization. Due to the geometry of the potential and the all-to-all coupling mediated by the cavity light field, all atoms on the even sites experience the same light-field interaction strength, as do the atoms on the odd sites. We determine the phase diagram for the square-shaped four-site realization with a mean-field ansatz. We consider a product state of coherent states as an ansatz
\begin{equation}
| \Psi\rangle = \left( |\psi_-\rangle|\psi_+\rangle \right)^{M/2} |\alpha \rangle
\end{equation}
where $\psi_{\pm}$ and $\alpha$ are $c$-numbers representing the amplitude of the coherent states. $\psi_+$ ($\psi_-$) corresponds to the condensate on the odd (even) sites. We compute the energy $E= \langle \Psi | H | \Psi \rangle$, define $\psi_\pm = \sqrt{\left(N_\mathrm{A} \pm \Delta\right)/M}$ with $N_\mathrm{A}$ the total particle number, and find
\begin{equation}
\label{eq:E}
E= \omega |\alpha|^2- 2 g \alpha_r \Delta  -2 J  \cos \theta \sqrt{N_\mathrm{A}^2-\Delta^2},
\end{equation}
where we denote $\alpha_r \equiv  \Re(\alpha)$. By minimizing the energy, we find the critical light-matter coupling strength to be $g_\mathrm{crit}= \sqrt{J \omega\cos(\theta)/N_\mathrm{A}}$.  For an open system with dissipation rate $\kappa$, we obtain a modified critical coupling strength,
\begin{equation}
\label{eq:gcrit}
g_\mathrm{crit}= \sqrt{\frac{J\cos\theta\left(\omega^2+\kappa^2 \right)}{N_\mathrm{A}\omega}}.
\end{equation}
Hence, the critical light-matter coupling strength can be significantly reduced via rotation of the setup, due to the dependence on $\cos\left(\theta\right)$. This dependence is the origin of the functionality as a rotation sensor, due to the dependence $\theta = \theta(\Omega)$. A change of $\Omega$ results in a change of the emitted light intensity in real time without any destructive measurement.

In the following, we determine the dynamics of the four-site realization via a numerical implementation of the open Truncated Wigner Approximation (TWA) method \cite{Cosme2019,Kongkhambut2022,Skulte2022}. The TWA is a semi-classical phase space method, which uses an ensemble of initial states sampled over the corresponding Wigner distribution to predict the quantum dynamics. For the cavity mode, we sample from a Wigner distribution corresponding to a coherent state with $\langle \alpha \rangle=0$. For the atoms on each site, we sample from a Wigner distribution of a coherent state with $\langle n_i\rangle=N_\mathrm{A}/M$.
Due to the dissipative nature of the system, we propagate the initial states via stochastic differential equations. The equations of motion are given by the Heisenberg-Langevin equations
 \begin{align}
 \frac{d\hat{b}_i}{dt}&=i [\hat{H},\hat{b}_i] \\
  \frac{d\hat{a}}{dt}&=i [\hat{H},\hat{a}] - \kappa \hat{a} +\xi
 \end{align}
 where $\kappa$ is the cavity dissipation rate and $\xi$ represents the noise associated with the dissipation. The noise fulfills the relation $\langle \xi^*(t')\xi(t) \rangle = \kappa \delta(t-t')$. In the following, we use experimentally realistic parameters \cite{Klinder2015}. In particular, we consider $^{87}$Rb atoms and choose a particle number of $N_\mathrm{A}=60\times 10^3$ and a recoil energy of $\omega_\mathrm{rec}=2\pi \times 3.5~\mathrm{kHz}$, which corresponds to a wavelength of $\lambda \approx 800~\mathrm{nm}$.
We further assume to be in or near the good cavity or recoil-resolved regime $\omega_\mathrm{rec} \approx \kappa = 2\pi \times 5~\mathrm{kHz}$. However, we want to stress that our proposed sensor is not limited to operation in this regime. Using a larger $\kappa$ is also feasible but with the trade-off that the number of photons detected in real time would be less than for smaller $\kappa$. For the tunnelling rate, we assume  $J=2\pi \times 2~\mathrm{kHz}$.\\


\begin{figure}[!htpb]
\centering
\includegraphics[width=1\columnwidth]{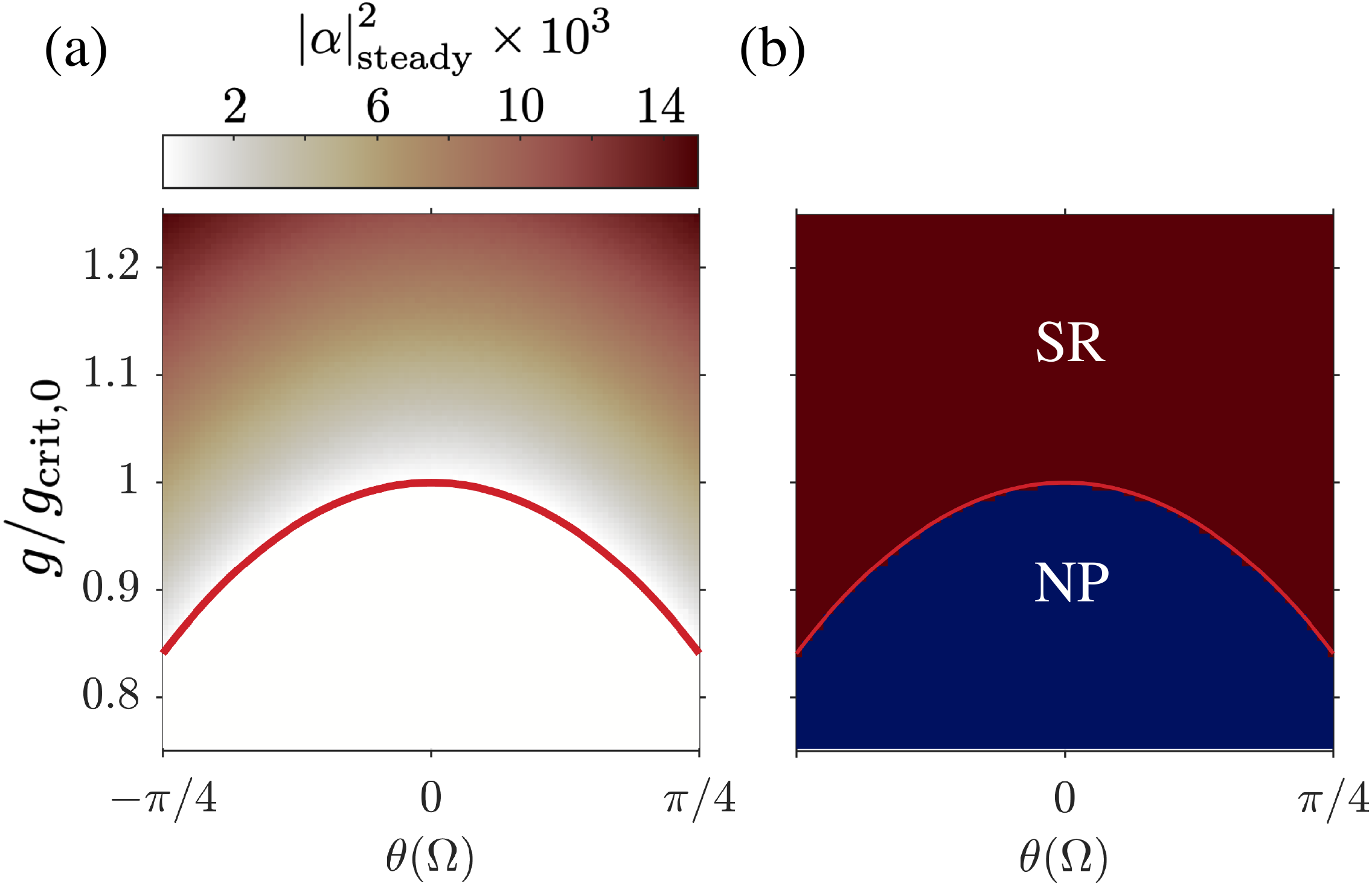}
\caption{(a) Steady state light field intensity $|\alpha|^2_\mathrm{steady}$ as a function of the rotation frequency and light-matter coupling strength. The light-matter coupling strength is rescaled by the critical value for the non-rotating case. For each data point, we average over $10^3$ TWA trajectories. The red curve corresponds to the analytically derived phase boundary in Eq.~(\ref{eq:gcrit}). (b) Phase diagram inferred from (a) using the criterion $|\alpha|^2_\mathrm{steady} > 10$ for the superradiant phase.} 
\label{fig:2} 
\end{figure}
In Fig.~\ref{fig:2}(a), we show the TWA results for the steady-state values of the photon number for varying gauge phase $\theta$ and light-matter coupling $g$ rescaled by the critical light-matter coupling for the system at rest, $g_{0,\mathrm{crit}}= g_\mathrm{crit}(\theta=0)$. The corresponding phase diagram is presented in Fig.~\ref{fig:2}(b). The red curve corresponds to the the analytical critical light-matter coupling strength in Eq.~(\ref{eq:gcrit}). We observe that the photon number strongly depends on the phase.
In Fig.~\ref{fig:2}(b), we distinguish between the normal phase (NP), in which there are no cavity photons and no imbalance between the population on the even and odd sites, and the SR phase, in which photons are scattered into the cavity and there is population imbalance between the even and odd sites. We note that for a fixed light-matter coupling the $\mathcal{Z}_2$ symmetry breaking phase transition between the NP and SR phase 
can be induced by changes in the gauge phase $\theta$ which originates from changes in the frequency of the rotation.
\begin{figure}[!htpb]
\label{fig:3} 
\centering
\includegraphics[width=1\columnwidth]{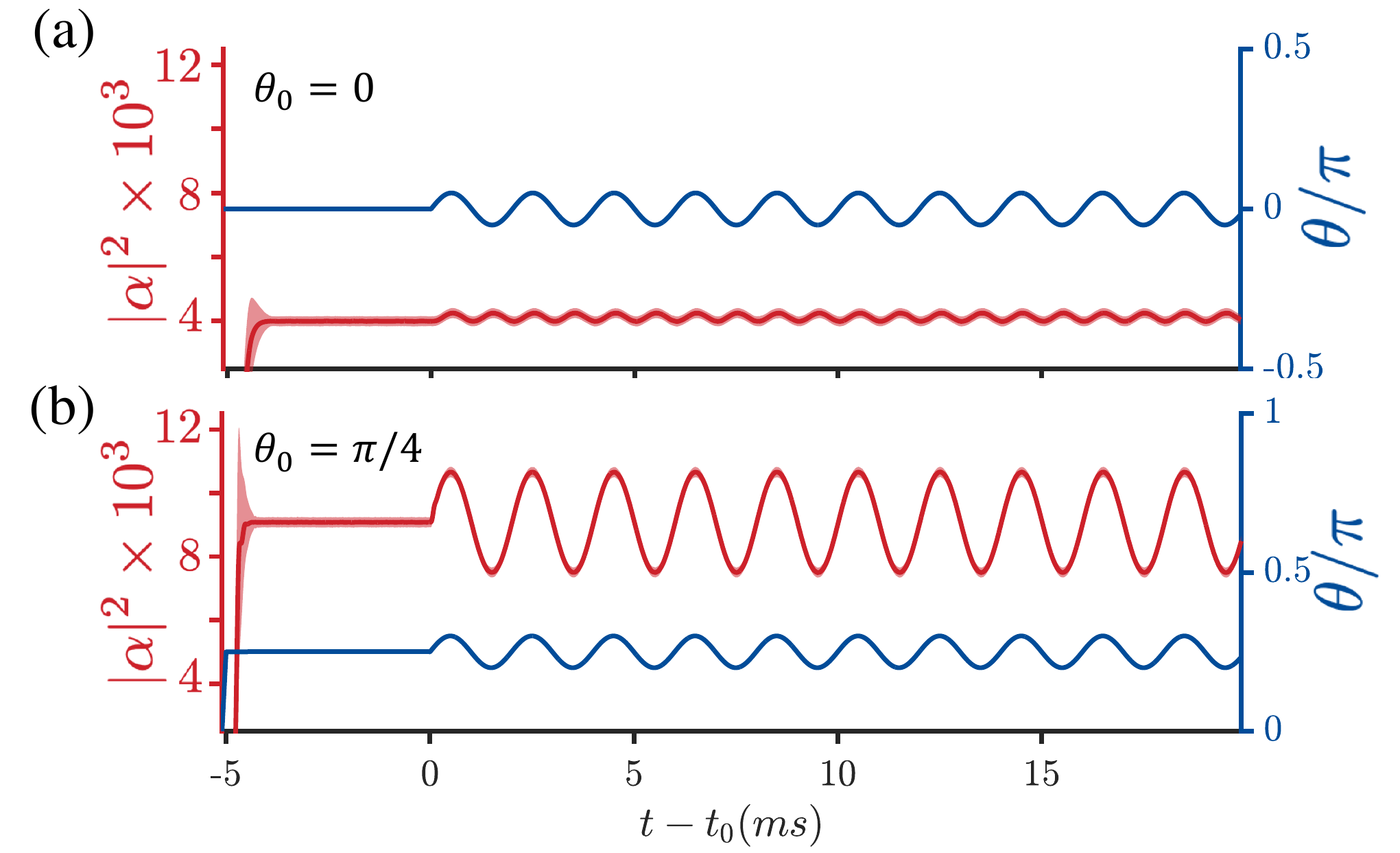}
\caption{(a) In blue, the applied phase $\theta$ modulated with a sinusoidal drive with strength $\delta_\theta = \pi/20$ and frequency $\omega_\mathrm{dr} = 2\pi \times 0.5~\mathrm{kHz}$. The dark red line shows the light field intensity derived from $10^3$ TWA trajectories. The light red shaded area shows the corresponding standard deviation due to quantum fluctuations within TWA. The system is initialized at rest before driving the system. (b) Same protocol as in (a) but the system is initialized with a phase of $\theta_0=\pi/4$. In both cases the amplitude change of $\Omega$ corresponds to  $\omega_\mathrm{rec}/(20 \pi n_\mathrm{s})=350~\mathrm{Hz}/n_\mathrm{s}$ and we choose $g=1.09 g_\mathrm{0,crit}$.} 
\label{fig:3} 
\end{figure}

In the following, we elaborate on dynamically sensing rotational frequencies in real time. The timescale in which changes of frequencies can be distinguished is roughly $1/J = 5 \times 10^{-4}~\mathrm{s}$. This makes the sensor an ideal candidate for inertial navigation, where quick dynamical changes in the the rotational frequency need to be detected. Further, we propose to increase the sensitivity of the sensor by using a bias rotation, i.e. rotating the system at a fixed frequency, so that the signal is the incremental change compared to the bias rotation. In Fig.~\ref{fig:3}(a), we present TWA results for the light field intensity in red and the corresponding standard deviation in the shaded area. The results are obtained using $10^3$ trajectories. The blue line corresponds to the phase $\theta$, which is now modulated in time, as an example for time-dependent behavior. We ramp up the light matter coupling to $g=1.09g_{0,\mathrm{crit}}$ within $1~\mathrm{ms}$ at the beginning of each run.
Initially the system is at rest, corresponding to $\theta=0$. After the time $t_0$ we change the phase via $\theta=\theta_0+\delta_\theta \sin(\omega_\mathrm{dr}t)$ to showcase the dynamical detection of changes in the rotational frequency. We choose $\omega_\mathrm{dr}=0.5~\mathrm{kHz}$ and $\delta_\theta=\pi/20$ corresponding to a frequency change of  $\delta_\Omega=\omega_\mathrm{rec} / 20 \pi n_\mathrm{s}=0.35/n_\mathrm{s}~\mathrm{kHz}$.\\
\indent As seen in Fig.~\ref{fig:2}(a) and Fig.~\ref{fig:3}(a) for $\theta=0$, any change in the phase will increase the light intensity. Due to this effect the applied phase change creates a frequency doubling in the light intensity. We emphasize that the magnitude of the relative change in the light field intensity is larger than the standard deviation or uncertainty due to quantum fluctuations.  Thus, we can reliably distinguish between the dynamically varying rotational frequencies. In Fig. 3(b), we use the same protocol as in Fig. 3(a) except that for the first $1~\mathrm{ms}$ ramp of the light-matter coupling, the phase $\theta_0$ is also ramped up from $0$ to $\pi/4$. This corresponds to rotating the sensor before the sensing starts. Fig.~\ref{fig:3}(b) shows that the overall photon number is increased for fixed light-matter coupling strength $g$. Furthermore the response of the system to the changes in the phase is more sensitive as inferred from the larger variations in the photon number compared to Fig.~\ref{fig:3}(a). For the specific choice of parameters considered here, the sensor must be rotated at approximately $\Omega_0=\omega_\mathrm{rec}/4 \pi n_\mathrm{s}=1.75/n_\mathrm{s}~\mathrm{kHz}$ to operate in this regime. \\ \indent We note that the change of the rotational frequency as depicted in Fig.~\ref{fig:3} can still be detected if we assume large particle number fluctuations with the width of $\sigma=0.1\times N_\mathrm{A}$ in each run of the measurement. This is shown in the supplemental material \cite{suppmat}. \\
\indent Before we conclude, we briefly discuss the influence of parameter choices on the sensitivity of the proposed rotational sensor. In order to prepare a dynamical and sensitive sensor we want to minimize the ratio $\Delta \Omega$ and $\tau$, where $\Delta \Omega$ is the smallest change in the rotational frequency that can be distinguished from noise and $\tau$ is the response time of the sensor. Generally speaking, $\tau$ is proportional to $1/J$. Hence, we want to maximize the tunnelling amplitude, while keeping it sufficiently small to avoid exciting atoms into higher bands. To increase the dynamical rotational sensitivity there are two routes one can take. The first is to increase the enclosed area by adding more sites to the array as discussed above, see Fig~\ref{fig:1}(c). The second choice is to change the platform for implementing the discrete model studied here. Specifically, the spacing between different sites could be increased to decrease $\omega_\mathrm{rec}$, which means that the system no longer operates in the optical regime but in the near infrared regime instead, for example. This suggests that this mechanism to detect changes in rotational frequencies is not limited to atom-cavity systems, but can be implemented in other light-matter coupled systems, with a different recoil energy of atomic sector and/or which allow to enclose even larger areas. This makes our approach tunable to the desired frequency domain in which the sensor is to operate. \\
\indent In conclusion, we have presented a new mechanism for dynamical rotational sensing. Our proposed system utilizes the artificial gauge field that neutral atoms experience due to external rotation, and combine it with the superradiant phase transition of an array of atoms coupled to a single mode cavity. This allows for a high precision measurement in real time. We further highlight the possibility that the sensitivity can be increased by increasing the enclosed area of the quantum sensor. A change of $\Delta \theta=\pi/400$ of the gauge field, corresponding to realistic parameters to $\approx 17.5~\mathrm{Hz}/M$ of the rotational frequency, can be measured and distinguished from the cavity noise, where $M$ denotes the total number of sites. We note that the influence of rotation on the superradiant phase transition can be observed in current state-of-the-art atom-cavity experiments \cite{Klinder2015,Landig2016} using an actuator giving a dynamical twist to the optical table. Our proposal puts forth to create quantum rotational sensors not only for static high precision frequency measurements but also for inertial measurements and navigation, where the ability to measure time-varying signals is crucial. \\ \indent 

\begin{acknowledgments}
We thank Andreas Hemmerich for useful discussions.
 This work was funded by the Deutsche Forschungsgemeinschaft (DFG, German Research Foundation) ``SFB-925" project 170620586 and the Cluster of Excellence ``Advanced Imaging of Matter" (EXC 2056), Project No. 390715994, and the UP System Balik PhD Program (OVPAA-BPhD-2021-04). J.S. acknowledges support from the German Academic Scholarship Foundation. 
\end{acknowledgments}

\bibliography{references_sensor}

\providecommand{\noopsort}[1]{}\providecommand{\singleletter}[1]{#1}%
\begin{thebibliography}{45}%
\makeatletter
\providecommand \@ifxundefined [1]{%
 \@ifx{#1\undefined}
}%
\providecommand \@ifnum [1]{%
 \ifnum #1\expandafter \@firstoftwo
 \else \expandafter \@secondoftwo
 \fi
}%
\providecommand \@ifx [1]{%
 \ifx #1\expandafter \@firstoftwo
 \else \expandafter \@secondoftwo
 \fi
}%
\providecommand \natexlab [1]{#1}%
\providecommand \enquote  [1]{``#1''}%
\providecommand \bibnamefont  [1]{#1}%
\providecommand \bibfnamefont [1]{#1}%
\providecommand \citenamefont [1]{#1}%
\providecommand \href@noop [0]{\@secondoftwo}%
\providecommand \href [0]{\begingroup \@sanitize@url \@href}%
\providecommand \@href[1]{\@@startlink{#1}\@@href}%
\providecommand \@@href[1]{\endgroup#1\@@endlink}%
\providecommand \@sanitize@url [0]{\catcode `\\12\catcode `\$12\catcode
  `\&12\catcode `\#12\catcode `\^12\catcode `\_12\catcode `\%12\relax}%
\providecommand \@@startlink[1]{}%
\providecommand \@@endlink[0]{}%
\providecommand \url  [0]{\begingroup\@sanitize@url \@url }%
\providecommand \@url [1]{\endgroup\@href {#1}{\urlprefix }}%
\providecommand \urlprefix  [0]{URL }%
\providecommand \Eprint [0]{\href }%
\providecommand \doibase [0]{https://doi.org/}%
\providecommand \selectlanguage [0]{\@gobble}%
\providecommand \bibinfo  [0]{\@secondoftwo}%
\providecommand \bibfield  [0]{\@secondoftwo}%
\providecommand \translation [1]{[#1]}%
\providecommand \BibitemOpen [0]{}%
\providecommand \bibitemStop [0]{}%
\providecommand \bibitemNoStop [0]{.\EOS\space}%
\providecommand \EOS [0]{\spacefactor3000\relax}%
\providecommand \BibitemShut  [1]{\csname bibitem#1\endcsname}%
\let\auto@bib@innerbib\@empty
\bibitem [{\citenamefont {Amico}\ \emph {et~al.}(2021)\citenamefont {Amico},
  \citenamefont {Boshier}, \citenamefont {Birkl}, \citenamefont {Minguzzi},
  \citenamefont {Miniatura}, \citenamefont {Kwek}, \citenamefont {Aghamalyan},
  \citenamefont {Ahufinger}, \citenamefont {Anderson}, \citenamefont {Andrei},
  \citenamefont {Arnold}, \citenamefont {Baker}, \citenamefont {Bell},
  \citenamefont {Bland}, \citenamefont {Brantut}, \citenamefont {Cassettari},
  \citenamefont {Chetcuti}, \citenamefont {Chevy}, \citenamefont {Citro},
  \citenamefont {De~Palo}, \citenamefont {Dumke}, \citenamefont {Edwards},
  \citenamefont {Folman}, \citenamefont {Fortagh}, \citenamefont {Gardiner},
  \citenamefont {Garraway}, \citenamefont {Gauthier}, \citenamefont {Günther},
  \citenamefont {Haug}, \citenamefont {Hufnagel}, \citenamefont {Keil},
  \citenamefont {Ireland}, \citenamefont {Lebrat}, \citenamefont {Li},
  \citenamefont {Longchambon}, \citenamefont {Mompart}, \citenamefont {Morsch},
  \citenamefont {Naldesi}, \citenamefont {Neely}, \citenamefont {Olshanii},
  \citenamefont {Orignac}, \citenamefont {Pandey}, \citenamefont
  {Pérez-Obiol}, \citenamefont {Perrin}, \citenamefont {Piroli}, \citenamefont
  {Polo}, \citenamefont {Pritchard}, \citenamefont {Proukakis}, \citenamefont
  {Rylands}, \citenamefont {Rubinsztein-Dunlop}, \citenamefont {Scazza},
  \citenamefont {Stringari}, \citenamefont {Tosto}, \citenamefont
  {Trombettoni}, \citenamefont {Victorin}, \citenamefont {Klitzing},
  \citenamefont {Wilkowski}, \citenamefont {Xhani},\ and\ \citenamefont
  {Yakimenko}}]{Amico2021}%
  \BibitemOpen
  \bibfield  {author} {\bibinfo {author} {\bibfnamefont {L.}~\bibnamefont
  {Amico}}, \bibinfo {author} {\bibfnamefont {M.}~\bibnamefont {Boshier}},
  \bibinfo {author} {\bibfnamefont {G.}~\bibnamefont {Birkl}}, \bibinfo
  {author} {\bibfnamefont {A.}~\bibnamefont {Minguzzi}}, \bibinfo {author}
  {\bibfnamefont {C.}~\bibnamefont {Miniatura}}, \bibinfo {author}
  {\bibfnamefont {L.-C.}\ \bibnamefont {Kwek}}, \bibinfo {author}
  {\bibfnamefont {D.}~\bibnamefont {Aghamalyan}}, \bibinfo {author}
  {\bibfnamefont {V.}~\bibnamefont {Ahufinger}}, \bibinfo {author}
  {\bibfnamefont {D.}~\bibnamefont {Anderson}}, \bibinfo {author}
  {\bibfnamefont {N.}~\bibnamefont {Andrei}}, \bibinfo {author} {\bibfnamefont
  {A.~S.}\ \bibnamefont {Arnold}}, \bibinfo {author} {\bibfnamefont
  {M.}~\bibnamefont {Baker}}, \bibinfo {author} {\bibfnamefont {T.~A.}\
  \bibnamefont {Bell}}, \bibinfo {author} {\bibfnamefont {T.}~\bibnamefont
  {Bland}}, \bibinfo {author} {\bibfnamefont {J.~P.}\ \bibnamefont {Brantut}},
  \bibinfo {author} {\bibfnamefont {D.}~\bibnamefont {Cassettari}}, \bibinfo
  {author} {\bibfnamefont {W.~J.}\ \bibnamefont {Chetcuti}}, \bibinfo {author}
  {\bibfnamefont {F.}~\bibnamefont {Chevy}}, \bibinfo {author} {\bibfnamefont
  {R.}~\bibnamefont {Citro}}, \bibinfo {author} {\bibfnamefont
  {S.}~\bibnamefont {De~Palo}}, \bibinfo {author} {\bibfnamefont
  {R.}~\bibnamefont {Dumke}}, \bibinfo {author} {\bibfnamefont
  {M.}~\bibnamefont {Edwards}}, \bibinfo {author} {\bibfnamefont
  {R.}~\bibnamefont {Folman}}, \bibinfo {author} {\bibfnamefont
  {J.}~\bibnamefont {Fortagh}}, \bibinfo {author} {\bibfnamefont {S.~A.}\
  \bibnamefont {Gardiner}}, \bibinfo {author} {\bibfnamefont {B.~M.}\
  \bibnamefont {Garraway}}, \bibinfo {author} {\bibfnamefont {G.}~\bibnamefont
  {Gauthier}}, \bibinfo {author} {\bibfnamefont {A.}~\bibnamefont {Günther}},
  \bibinfo {author} {\bibfnamefont {T.}~\bibnamefont {Haug}}, \bibinfo {author}
  {\bibfnamefont {C.}~\bibnamefont {Hufnagel}}, \bibinfo {author}
  {\bibfnamefont {M.}~\bibnamefont {Keil}}, \bibinfo {author} {\bibfnamefont
  {P.}~\bibnamefont {Ireland}}, \bibinfo {author} {\bibfnamefont
  {M.}~\bibnamefont {Lebrat}}, \bibinfo {author} {\bibfnamefont
  {W.}~\bibnamefont {Li}}, \bibinfo {author} {\bibfnamefont {L.}~\bibnamefont
  {Longchambon}}, \bibinfo {author} {\bibfnamefont {J.}~\bibnamefont
  {Mompart}}, \bibinfo {author} {\bibfnamefont {O.}~\bibnamefont {Morsch}},
  \bibinfo {author} {\bibfnamefont {P.}~\bibnamefont {Naldesi}}, \bibinfo
  {author} {\bibfnamefont {T.~W.}\ \bibnamefont {Neely}}, \bibinfo {author}
  {\bibfnamefont {M.}~\bibnamefont {Olshanii}}, \bibinfo {author}
  {\bibfnamefont {E.}~\bibnamefont {Orignac}}, \bibinfo {author} {\bibfnamefont
  {S.}~\bibnamefont {Pandey}}, \bibinfo {author} {\bibfnamefont
  {A.}~\bibnamefont {Pérez-Obiol}}, \bibinfo {author} {\bibfnamefont
  {H.}~\bibnamefont {Perrin}}, \bibinfo {author} {\bibfnamefont
  {L.}~\bibnamefont {Piroli}}, \bibinfo {author} {\bibfnamefont
  {J.}~\bibnamefont {Polo}}, \bibinfo {author} {\bibfnamefont {A.~L.}\
  \bibnamefont {Pritchard}}, \bibinfo {author} {\bibfnamefont {N.~P.}\
  \bibnamefont {Proukakis}}, \bibinfo {author} {\bibfnamefont {C.}~\bibnamefont
  {Rylands}}, \bibinfo {author} {\bibfnamefont {H.}~\bibnamefont
  {Rubinsztein-Dunlop}}, \bibinfo {author} {\bibfnamefont {F.}~\bibnamefont
  {Scazza}}, \bibinfo {author} {\bibfnamefont {S.}~\bibnamefont {Stringari}},
  \bibinfo {author} {\bibfnamefont {F.}~\bibnamefont {Tosto}}, \bibinfo
  {author} {\bibfnamefont {A.}~\bibnamefont {Trombettoni}}, \bibinfo {author}
  {\bibfnamefont {N.}~\bibnamefont {Victorin}}, \bibinfo {author}
  {\bibfnamefont {W.~v.}\ \bibnamefont {Klitzing}}, \bibinfo {author}
  {\bibfnamefont {D.}~\bibnamefont {Wilkowski}}, \bibinfo {author}
  {\bibfnamefont {K.}~\bibnamefont {Xhani}},\ and\ \bibinfo {author}
  {\bibfnamefont {A.}~\bibnamefont {Yakimenko}},\ }\bibfield  {title} {\bibinfo
  {title} {Roadmap on atomtronics: State of the art and perspective},\ }\href
  {https://doi.org/10.1116/5.0026178} {\bibfield  {journal} {\bibinfo
  {journal} {AVS Quantum Science}\ }\textbf {\bibinfo {volume} {3}},\ \bibinfo
  {pages} {039201} (\bibinfo {year} {2021})}\BibitemShut {NoStop}%
\bibitem [{\citenamefont {Lef{\`e}vre}(1993)}]{lefevre1994}%
  \BibitemOpen
  \bibfield  {author} {\bibinfo {author} {\bibfnamefont {H.}~\bibnamefont
  {Lef{\`e}vre}},\ }\href@noop {} {\emph {\bibinfo {title} {The Fiber-optic
  Gyroscope}}},\ Artech House optoelectronics library\ (\bibinfo  {publisher}
  {Artech House},\ \bibinfo {year} {1993})\BibitemShut {NoStop}%
\bibitem [{\citenamefont {Durfee}\ \emph {et~al.}(2006)\citenamefont {Durfee},
  \citenamefont {Shaham},\ and\ \citenamefont {Kasevich}}]{Durfee2006}%
  \BibitemOpen
  \bibfield  {author} {\bibinfo {author} {\bibfnamefont {D.~S.}\ \bibnamefont
  {Durfee}}, \bibinfo {author} {\bibfnamefont {Y.~K.}\ \bibnamefont {Shaham}},\
  and\ \bibinfo {author} {\bibfnamefont {M.~A.}\ \bibnamefont {Kasevich}},\
  }\bibfield  {title} {\bibinfo {title} {Long-term stability of an
  area-reversible atom-interferometer sagnac gyroscope},\ }\href
  {https://doi.org/10.1103/PhysRevLett.97.240801} {\bibfield  {journal}
  {\bibinfo  {journal} {Phys. Rev. Lett.}\ }\textbf {\bibinfo {volume} {97}},\
  \bibinfo {pages} {240801} (\bibinfo {year} {2006})}\BibitemShut {NoStop}%
\bibitem [{\citenamefont {Burke}\ and\ \citenamefont
  {Sackett}(2009)}]{Burke2009}%
  \BibitemOpen
  \bibfield  {author} {\bibinfo {author} {\bibfnamefont {J.~H.~T.}\
  \bibnamefont {Burke}}\ and\ \bibinfo {author} {\bibfnamefont {C.~A.}\
  \bibnamefont {Sackett}},\ }\bibfield  {title} {\bibinfo {title} {Scalable
  bose-einstein-condensate sagnac interferometer in a linear trap},\ }\href
  {https://doi.org/10.1103/PhysRevA.80.061603} {\bibfield  {journal} {\bibinfo
  {journal} {Phys. Rev. A}\ }\textbf {\bibinfo {volume} {80}},\ \bibinfo
  {pages} {061603} (\bibinfo {year} {2009})}\BibitemShut {NoStop}%
\bibitem [{\citenamefont {Hamilton}\ \emph {et~al.}(2015)\citenamefont
  {Hamilton}, \citenamefont {Jaffe}, \citenamefont {Brown}, \citenamefont
  {Maisenbacher}, \citenamefont {Estey},\ and\ \citenamefont
  {M\"uller}}]{Hamilton2015}%
  \BibitemOpen
  \bibfield  {author} {\bibinfo {author} {\bibfnamefont {P.}~\bibnamefont
  {Hamilton}}, \bibinfo {author} {\bibfnamefont {M.}~\bibnamefont {Jaffe}},
  \bibinfo {author} {\bibfnamefont {J.~M.}\ \bibnamefont {Brown}}, \bibinfo
  {author} {\bibfnamefont {L.}~\bibnamefont {Maisenbacher}}, \bibinfo {author}
  {\bibfnamefont {B.}~\bibnamefont {Estey}},\ and\ \bibinfo {author}
  {\bibfnamefont {H.}~\bibnamefont {M\"uller}},\ }\bibfield  {title} {\bibinfo
  {title} {Atom interferometry in an optical cavity},\ }\href
  {https://doi.org/10.1103/PhysRevLett.114.100405} {\bibfield  {journal}
  {\bibinfo  {journal} {Phys. Rev. Lett.}\ }\textbf {\bibinfo {volume} {114}},\
  \bibinfo {pages} {100405} (\bibinfo {year} {2015})}\BibitemShut {NoStop}%
\bibitem [{\citenamefont {Bongs}\ \emph {et~al.}(2019)\citenamefont {Bongs},
  \citenamefont {Holynski}, \citenamefont {Vovrosh}, \citenamefont {Bouyer},
  \citenamefont {Condon}, \citenamefont {Rasel}, \citenamefont {Schubert},
  \citenamefont {Schleich},\ and\ \citenamefont {Roura}}]{Bongs2019}%
  \BibitemOpen
  \bibfield  {author} {\bibinfo {author} {\bibfnamefont {K.}~\bibnamefont
  {Bongs}}, \bibinfo {author} {\bibfnamefont {M.}~\bibnamefont {Holynski}},
  \bibinfo {author} {\bibfnamefont {J.}~\bibnamefont {Vovrosh}}, \bibinfo
  {author} {\bibfnamefont {P.}~\bibnamefont {Bouyer}}, \bibinfo {author}
  {\bibfnamefont {G.}~\bibnamefont {Condon}}, \bibinfo {author} {\bibfnamefont
  {E.}~\bibnamefont {Rasel}}, \bibinfo {author} {\bibfnamefont
  {C.}~\bibnamefont {Schubert}}, \bibinfo {author} {\bibfnamefont {W.~P.}\
  \bibnamefont {Schleich}},\ and\ \bibinfo {author} {\bibfnamefont
  {A.}~\bibnamefont {Roura}},\ }\bibfield  {title} {\bibinfo {title} {Taking
  atom interferometric quantum sensors from the laboratory to real-world
  applications},\ }\href {https://doi.org/10.1038/s42254-019-0117-4} {\bibfield
   {journal} {\bibinfo  {journal} {Nature Reviews Physics}\ }\textbf {\bibinfo
  {volume} {1}},\ \bibinfo {pages} {731} (\bibinfo {year} {2019})}\BibitemShut
  {NoStop}%
\bibitem [{\citenamefont {Moan}\ \emph {et~al.}(2020)\citenamefont {Moan},
  \citenamefont {Horne}, \citenamefont {Arpornthip}, \citenamefont {Luo},
  \citenamefont {Fallon}, \citenamefont {Berl},\ and\ \citenamefont
  {Sackett}}]{Moan2020}%
  \BibitemOpen
  \bibfield  {author} {\bibinfo {author} {\bibfnamefont {E.~R.}\ \bibnamefont
  {Moan}}, \bibinfo {author} {\bibfnamefont {R.~A.}\ \bibnamefont {Horne}},
  \bibinfo {author} {\bibfnamefont {T.}~\bibnamefont {Arpornthip}}, \bibinfo
  {author} {\bibfnamefont {Z.}~\bibnamefont {Luo}}, \bibinfo {author}
  {\bibfnamefont {A.~J.}\ \bibnamefont {Fallon}}, \bibinfo {author}
  {\bibfnamefont {S.~J.}\ \bibnamefont {Berl}},\ and\ \bibinfo {author}
  {\bibfnamefont {C.~A.}\ \bibnamefont {Sackett}},\ }\bibfield  {title}
  {\bibinfo {title} {Quantum rotation sensing with dual sagnac interferometers
  in an atom-optical waveguide},\ }\href
  {https://doi.org/10.1103/PhysRevLett.124.120403} {\bibfield  {journal}
  {\bibinfo  {journal} {Phys. Rev. Lett.}\ }\textbf {\bibinfo {volume} {124}},\
  \bibinfo {pages} {120403} (\bibinfo {year} {2020})}\BibitemShut {NoStop}%
\bibitem [{\citenamefont {Woffinden}\ \emph {et~al.}(2022)\citenamefont
  {Woffinden}, \citenamefont {Groszek}, \citenamefont {Gauthier}, \citenamefont
  {Mommers}, \citenamefont {Bromley}, \citenamefont {Haine}, \citenamefont
  {Rubinsztein-Dunlop}, \citenamefont {Davis}, \citenamefont {Neely},\ and\
  \citenamefont {Baker}}]{Woffinden2022}%
  \BibitemOpen
  \bibfield  {author} {\bibinfo {author} {\bibfnamefont {C.~W.}\ \bibnamefont
  {Woffinden}}, \bibinfo {author} {\bibfnamefont {A.~J.}\ \bibnamefont
  {Groszek}}, \bibinfo {author} {\bibfnamefont {G.}~\bibnamefont {Gauthier}},
  \bibinfo {author} {\bibfnamefont {B.~J.}\ \bibnamefont {Mommers}}, \bibinfo
  {author} {\bibfnamefont {M.~W.~J.}\ \bibnamefont {Bromley}}, \bibinfo
  {author} {\bibfnamefont {S.~A.}\ \bibnamefont {Haine}}, \bibinfo {author}
  {\bibfnamefont {H.}~\bibnamefont {Rubinsztein-Dunlop}}, \bibinfo {author}
  {\bibfnamefont {M.~J.}\ \bibnamefont {Davis}}, \bibinfo {author}
  {\bibfnamefont {T.~W.}\ \bibnamefont {Neely}},\ and\ \bibinfo {author}
  {\bibfnamefont {M.}~\bibnamefont {Baker}},\ }\href
  {https://doi.org/10.48550/ARXIV.2212.11617} {\bibinfo {title} {Viability of
  rotation sensing using phonon interferometry in bose-einstein condensates}}
  (\bibinfo {year} {2022})\BibitemShut {NoStop}%
\bibitem [{\citenamefont {Bouchendira}\ \emph {et~al.}(2011)\citenamefont
  {Bouchendira}, \citenamefont {Clad\'e}, \citenamefont {Guellati-Kh\'elifa},
  \citenamefont {Nez},\ and\ \citenamefont {Biraben}}]{Bouchendira2011}%
  \BibitemOpen
  \bibfield  {author} {\bibinfo {author} {\bibfnamefont {R.}~\bibnamefont
  {Bouchendira}}, \bibinfo {author} {\bibfnamefont {P.}~\bibnamefont
  {Clad\'e}}, \bibinfo {author} {\bibfnamefont {S.}~\bibnamefont
  {Guellati-Kh\'elifa}}, \bibinfo {author} {\bibfnamefont {F.}~\bibnamefont
  {Nez}},\ and\ \bibinfo {author} {\bibfnamefont {F.}~\bibnamefont {Biraben}},\
  }\bibfield  {title} {\bibinfo {title} {New determination of the fine
  structure constant and test of the quantum electrodynamics},\ }\href
  {https://doi.org/10.1103/PhysRevLett.106.080801} {\bibfield  {journal}
  {\bibinfo  {journal} {Phys. Rev. Lett.}\ }\textbf {\bibinfo {volume} {106}},\
  \bibinfo {pages} {080801} (\bibinfo {year} {2011})}\BibitemShut {NoStop}%
\bibitem [{\citenamefont {Fixler}\ \emph {et~al.}(2007)\citenamefont {Fixler},
  \citenamefont {Foster}, \citenamefont {McGuirk},\ and\ \citenamefont
  {Kasevich}}]{Fixler2007}%
  \BibitemOpen
  \bibfield  {author} {\bibinfo {author} {\bibfnamefont {J.~B.}\ \bibnamefont
  {Fixler}}, \bibinfo {author} {\bibfnamefont {G.~T.}\ \bibnamefont {Foster}},
  \bibinfo {author} {\bibfnamefont {J.~M.}\ \bibnamefont {McGuirk}},\ and\
  \bibinfo {author} {\bibfnamefont {M.~A.}\ \bibnamefont {Kasevich}},\
  }\bibfield  {title} {\bibinfo {title} {Atom interferometer measurement of the
  newtonian constant of gravity},\ }\href
  {https://doi.org/10.1126/science.1135459} {\bibfield  {journal} {\bibinfo
  {journal} {Science}\ }\textbf {\bibinfo {volume} {315}},\ \bibinfo {pages}
  {74} (\bibinfo {year} {2007})}\BibitemShut {NoStop}%
\bibitem [{\citenamefont {Parker}\ \emph {et~al.}(2018)\citenamefont {Parker},
  \citenamefont {Yu}, \citenamefont {Zhong}, \citenamefont {Estey},\ and\
  \citenamefont {Müller}}]{Parker2018}%
  \BibitemOpen
  \bibfield  {author} {\bibinfo {author} {\bibfnamefont {R.~H.}\ \bibnamefont
  {Parker}}, \bibinfo {author} {\bibfnamefont {C.}~\bibnamefont {Yu}}, \bibinfo
  {author} {\bibfnamefont {W.}~\bibnamefont {Zhong}}, \bibinfo {author}
  {\bibfnamefont {B.}~\bibnamefont {Estey}},\ and\ \bibinfo {author}
  {\bibfnamefont {H.}~\bibnamefont {Müller}},\ }\bibfield  {title} {\bibinfo
  {title} {Measurement of the fine-structure constant as a test of the standard
  model},\ }\href {https://doi.org/10.1126/science.aap7706} {\bibfield
  {journal} {\bibinfo  {journal} {Science}\ }\textbf {\bibinfo {volume}
  {360}},\ \bibinfo {pages} {191} (\bibinfo {year} {2018})}\BibitemShut
  {NoStop}%
\bibitem [{\citenamefont {Rosi}\ \emph {et~al.}(2014)\citenamefont {Rosi},
  \citenamefont {Sorrentino}, \citenamefont {Cacciapuoti}, \citenamefont
  {Prevedelli},\ and\ \citenamefont {Tino}}]{Rosi2014}%
  \BibitemOpen
  \bibfield  {author} {\bibinfo {author} {\bibfnamefont {G.}~\bibnamefont
  {Rosi}}, \bibinfo {author} {\bibfnamefont {F.}~\bibnamefont {Sorrentino}},
  \bibinfo {author} {\bibfnamefont {L.}~\bibnamefont {Cacciapuoti}}, \bibinfo
  {author} {\bibfnamefont {M.}~\bibnamefont {Prevedelli}},\ and\ \bibinfo
  {author} {\bibfnamefont {G.~M.}\ \bibnamefont {Tino}},\ }\bibfield  {title}
  {\bibinfo {title} {Precision measurement of the newtonian gravitational
  constant using cold atoms},\ }\href {https://doi.org/10.1038/nature13433}
  {\bibfield  {journal} {\bibinfo  {journal} {Nature}\ }\textbf {\bibinfo
  {volume} {510}},\ \bibinfo {pages} {518} (\bibinfo {year}
  {2014})}\BibitemShut {NoStop}%
\bibitem [{\citenamefont {Freier}\ \emph {et~al.}(2016)\citenamefont {Freier},
  \citenamefont {Hauth}, \citenamefont {Schkolnik}, \citenamefont {Leykauf},
  \citenamefont {Schilling}, \citenamefont {Wziontek}, \citenamefont
  {Scherneck}, \citenamefont {Müller},\ and\ \citenamefont
  {Peters}}]{Freier2016}%
  \BibitemOpen
  \bibfield  {author} {\bibinfo {author} {\bibfnamefont {C.}~\bibnamefont
  {Freier}}, \bibinfo {author} {\bibfnamefont {M.}~\bibnamefont {Hauth}},
  \bibinfo {author} {\bibfnamefont {V.}~\bibnamefont {Schkolnik}}, \bibinfo
  {author} {\bibfnamefont {B.}~\bibnamefont {Leykauf}}, \bibinfo {author}
  {\bibfnamefont {M.}~\bibnamefont {Schilling}}, \bibinfo {author}
  {\bibfnamefont {H.}~\bibnamefont {Wziontek}}, \bibinfo {author}
  {\bibfnamefont {H.-G.}\ \bibnamefont {Scherneck}}, \bibinfo {author}
  {\bibfnamefont {J.}~\bibnamefont {Müller}},\ and\ \bibinfo {author}
  {\bibfnamefont {A.}~\bibnamefont {Peters}},\ }\bibfield  {title} {\bibinfo
  {title} {Mobile quantum gravity sensor with unprecedented stability},\ }\href
  {https://doi.org/10.1088/1742-6596/723/1/012050} {\bibfield  {journal}
  {\bibinfo  {journal} {Journal of Physics: Conference Series}\ }\textbf
  {\bibinfo {volume} {723}},\ \bibinfo {pages} {012050} (\bibinfo {year}
  {2016})}\BibitemShut {NoStop}%
\bibitem [{\citenamefont {Karcher}\ \emph {et~al.}(2018)\citenamefont
  {Karcher}, \citenamefont {Imanaliev}, \citenamefont {Merlet},\ and\
  \citenamefont {Pereira Dos~Santos}}]{Karcher2018}%
  \BibitemOpen
  \bibfield  {author} {\bibinfo {author} {\bibfnamefont {R.}~\bibnamefont
  {Karcher}}, \bibinfo {author} {\bibfnamefont {A.}~\bibnamefont {Imanaliev}},
  \bibinfo {author} {\bibfnamefont {S.}~\bibnamefont {Merlet}},\ and\ \bibinfo
  {author} {\bibfnamefont {F.}~\bibnamefont {Pereira Dos~Santos}},\ }\bibfield
  {title} {\bibinfo {title} {Improving the accuracy of atom interferometers
  with ultracold sources},\ }\href {https://doi.org/10.1088/1367-2630/aaf07d}
  {\bibfield  {journal} {\bibinfo  {journal} {New Journal of Physics}\ }\textbf
  {\bibinfo {volume} {20}},\ \bibinfo {pages} {113041} (\bibinfo {year}
  {2018})}\BibitemShut {NoStop}%
\bibitem [{\citenamefont {Grewal}\ \emph {et~al.}(2020)\citenamefont {Grewal},
  \citenamefont {Andrews},\ and\ \citenamefont {Bartone}}]{Grewal2020}%
  \BibitemOpen
  \bibfield  {author} {\bibinfo {author} {\bibfnamefont {M.}~\bibnamefont
  {Grewal}}, \bibinfo {author} {\bibfnamefont {A.}~\bibnamefont {Andrews}},\
  and\ \bibinfo {author} {\bibfnamefont {C.}~\bibnamefont {Bartone}},\
  }\href@noop {} {\emph {\bibinfo {title} {Global Navigation Satellite Systems,
  Inertial Navigation, and Integration}}}\ (\bibinfo  {publisher} {Wiley},\
  \bibinfo {year} {2020})\BibitemShut {NoStop}%
\bibitem [{\citenamefont {Geiger}\ \emph {et~al.}(2020)\citenamefont {Geiger},
  \citenamefont {Landragin}, \citenamefont {Merlet},\ and\ \citenamefont
  {Pereira Dos~Santos}}]{Geiger2020}%
  \BibitemOpen
  \bibfield  {author} {\bibinfo {author} {\bibfnamefont {R.}~\bibnamefont
  {Geiger}}, \bibinfo {author} {\bibfnamefont {A.}~\bibnamefont {Landragin}},
  \bibinfo {author} {\bibfnamefont {S.}~\bibnamefont {Merlet}},\ and\ \bibinfo
  {author} {\bibfnamefont {F.}~\bibnamefont {Pereira Dos~Santos}},\ }\bibfield
  {title} {\bibinfo {title} {High-accuracy inertial measurements with cold-atom
  sensors},\ }\href {https://doi.org/10.1116/5.0009093} {\bibfield  {journal}
  {\bibinfo  {journal} {AVS Quantum Science}\ }\textbf {\bibinfo {volume}
  {2}},\ \bibinfo {pages} {024702} (\bibinfo {year} {2020})}\BibitemShut
  {NoStop}%
\bibitem [{\citenamefont {Savoie}\ \emph {et~al.}(2018)\citenamefont {Savoie},
  \citenamefont {Altorio}, \citenamefont {Fang}, \citenamefont {Sidorenkov},
  \citenamefont {Geiger},\ and\ \citenamefont {Landragin}}]{Savoie2018}%
  \BibitemOpen
  \bibfield  {author} {\bibinfo {author} {\bibfnamefont {D.}~\bibnamefont
  {Savoie}}, \bibinfo {author} {\bibfnamefont {M.}~\bibnamefont {Altorio}},
  \bibinfo {author} {\bibfnamefont {B.}~\bibnamefont {Fang}}, \bibinfo {author}
  {\bibfnamefont {L.~A.}\ \bibnamefont {Sidorenkov}}, \bibinfo {author}
  {\bibfnamefont {R.}~\bibnamefont {Geiger}},\ and\ \bibinfo {author}
  {\bibfnamefont {A.}~\bibnamefont {Landragin}},\ }\bibfield  {title} {\bibinfo
  {title} {Interleaved atom interferometry for high-sensitivity inertial
  measurements},\ }\href {https://doi.org/10.1126/sciadv.aau7948} {\bibfield
  {journal} {\bibinfo  {journal} {Science Advances}\ }\textbf {\bibinfo
  {volume} {4}},\ \bibinfo {pages} {eaau7948} (\bibinfo {year}
  {2018})}\BibitemShut {NoStop}%
\bibitem [{\citenamefont {JEekeli}(2005)}]{JEKELI2005}%
  \BibitemOpen
  \bibfield  {author} {\bibinfo {author} {\bibfnamefont {C.}~\bibnamefont
  {JEekeli}},\ }\bibfield  {title} {\bibinfo {title} {Navigation error analysis
  of atom interferometer inertial sensor},\ }\href
  {https://doi.org/https://doi.org/10.1002/j.2161-4296.2005.tb01726.x}
  {\bibfield  {journal} {\bibinfo  {journal} {NAVIGATION}\ }\textbf {\bibinfo
  {volume} {52}},\ \bibinfo {pages} {1} (\bibinfo {year} {2005})}\BibitemShut
  {NoStop}%
\bibitem [{\citenamefont {Bhat}\ \emph {et~al.}(2006)\citenamefont {Bhat},
  \citenamefont {Holland},\ and\ \citenamefont {Carr}}]{Bhat2006}%
  \BibitemOpen
  \bibfield  {author} {\bibinfo {author} {\bibfnamefont {R.}~\bibnamefont
  {Bhat}}, \bibinfo {author} {\bibfnamefont {M.~J.}\ \bibnamefont {Holland}},\
  and\ \bibinfo {author} {\bibfnamefont {L.~D.}\ \bibnamefont {Carr}},\
  }\bibfield  {title} {\bibinfo {title} {Bose-einstein condensates in rotating
  lattices},\ }\href {https://doi.org/10.1103/PhysRevLett.96.060405} {\bibfield
   {journal} {\bibinfo  {journal} {Phys. Rev. Lett.}\ }\textbf {\bibinfo
  {volume} {96}},\ \bibinfo {pages} {060405} (\bibinfo {year}
  {2006})}\BibitemShut {NoStop}%
\bibitem [{\citenamefont {Rey}\ \emph {et~al.}(2007)\citenamefont {Rey},
  \citenamefont {Burnett}, \citenamefont {Satija},\ and\ \citenamefont
  {Clark}}]{Rey2007}%
  \BibitemOpen
  \bibfield  {author} {\bibinfo {author} {\bibfnamefont {A.~M.}\ \bibnamefont
  {Rey}}, \bibinfo {author} {\bibfnamefont {K.}~\bibnamefont {Burnett}},
  \bibinfo {author} {\bibfnamefont {I.~I.}\ \bibnamefont {Satija}},\ and\
  \bibinfo {author} {\bibfnamefont {C.~W.}\ \bibnamefont {Clark}},\ }\bibfield
  {title} {\bibinfo {title} {Entanglement and the mott transition in a rotating
  bosonic ring lattice},\ }\href {https://doi.org/10.1103/PhysRevA.75.063616}
  {\bibfield  {journal} {\bibinfo  {journal} {Phys. Rev. A}\ }\textbf {\bibinfo
  {volume} {75}},\ \bibinfo {pages} {063616} (\bibinfo {year}
  {2007})}\BibitemShut {NoStop}%
\bibitem [{\citenamefont {Nunnenkamp}\ \emph {et~al.}(2008)\citenamefont
  {Nunnenkamp}, \citenamefont {Rey},\ and\ \citenamefont
  {Burnett}}]{Nunnenkamp2008}%
  \BibitemOpen
  \bibfield  {author} {\bibinfo {author} {\bibfnamefont {A.}~\bibnamefont
  {Nunnenkamp}}, \bibinfo {author} {\bibfnamefont {A.~M.}\ \bibnamefont
  {Rey}},\ and\ \bibinfo {author} {\bibfnamefont {K.}~\bibnamefont {Burnett}},\
  }\bibfield  {title} {\bibinfo {title} {Generation of macroscopic
  superposition states in ring superlattices},\ }\href
  {https://doi.org/10.1103/PhysRevA.77.023622} {\bibfield  {journal} {\bibinfo
  {journal} {Phys. Rev. A}\ }\textbf {\bibinfo {volume} {77}},\ \bibinfo
  {pages} {023622} (\bibinfo {year} {2008})}\BibitemShut {NoStop}%
\bibitem [{\citenamefont {Hallwood}\ \emph {et~al.}(2006)\citenamefont
  {Hallwood}, \citenamefont {Burnett},\ and\ \citenamefont
  {Dunningham}}]{Hallwood2006}%
  \BibitemOpen
  \bibfield  {author} {\bibinfo {author} {\bibfnamefont {D.~W.}\ \bibnamefont
  {Hallwood}}, \bibinfo {author} {\bibfnamefont {K.}~\bibnamefont {Burnett}},\
  and\ \bibinfo {author} {\bibfnamefont {J.}~\bibnamefont {Dunningham}},\
  }\bibfield  {title} {\bibinfo {title} {Macroscopic superpositions of
  superfluid flows},\ }\href {https://doi.org/10.1088/1367-2630/8/9/180}
  {\bibfield  {journal} {\bibinfo  {journal} {New Journal of Physics}\ }\textbf
  {\bibinfo {volume} {8}},\ \bibinfo {pages} {180} (\bibinfo {year}
  {2006})}\BibitemShut {NoStop}%
\bibitem [{\citenamefont {Jiang}\ \emph {et~al.}(2022)\citenamefont {Jiang},
  \citenamefont {Zeng}, \citenamefont {Qin}, \citenamefont {Gong},\ and\
  \citenamefont {Fu}}]{Jiang2022}%
  \BibitemOpen
  \bibfield  {author} {\bibinfo {author} {\bibfnamefont {C.}~\bibnamefont
  {Jiang}}, \bibinfo {author} {\bibfnamefont {Y.}~\bibnamefont {Zeng}},
  \bibinfo {author} {\bibfnamefont {Q.}~\bibnamefont {Qin}}, \bibinfo {author}
  {\bibfnamefont {Z.}~\bibnamefont {Gong}},\ and\ \bibinfo {author}
  {\bibfnamefont {H.}~\bibnamefont {Fu}},\ }\href
  {https://doi.org/10.48550/ARXIV.2206.09318} {\bibinfo {title} {Quantum
  sensing of rotation velocity based on bose-hubbard model}} (\bibinfo {year}
  {2022})\BibitemShut {NoStop}%
\bibitem [{\citenamefont {Ritsch}\ \emph {et~al.}(2013)\citenamefont {Ritsch},
  \citenamefont {Domokos}, \citenamefont {Brennecke},\ and\ \citenamefont
  {Esslinger}}]{Ritsch2013}%
  \BibitemOpen
  \bibfield  {author} {\bibinfo {author} {\bibfnamefont {H.}~\bibnamefont
  {Ritsch}}, \bibinfo {author} {\bibfnamefont {P.}~\bibnamefont {Domokos}},
  \bibinfo {author} {\bibfnamefont {F.}~\bibnamefont {Brennecke}},\ and\
  \bibinfo {author} {\bibfnamefont {T.}~\bibnamefont {Esslinger}},\ }\bibfield
  {title} {\bibinfo {title} {Cold atoms in cavity-generated dynamical optical
  potentials},\ }\href {https://doi.org/10.1103/RevModPhys.85.553} {\bibfield
  {journal} {\bibinfo  {journal} {Rev. Mod. Phys.}\ }\textbf {\bibinfo {volume}
  {85}},\ \bibinfo {pages} {553} (\bibinfo {year} {2013})}\BibitemShut
  {NoStop}%
\bibitem [{\citenamefont {Mivehvar}\ \emph {et~al.}(2021)\citenamefont
  {Mivehvar}, \citenamefont {Piazza}, \citenamefont {Donner},\ and\
  \citenamefont {Ritsch}}]{Mivehvar2021}%
  \BibitemOpen
  \bibfield  {author} {\bibinfo {author} {\bibfnamefont {F.}~\bibnamefont
  {Mivehvar}}, \bibinfo {author} {\bibfnamefont {F.}~\bibnamefont {Piazza}},
  \bibinfo {author} {\bibfnamefont {T.}~\bibnamefont {Donner}},\ and\ \bibinfo
  {author} {\bibfnamefont {H.}~\bibnamefont {Ritsch}},\ }\bibfield  {title}
  {\bibinfo {title} {Cavity qed with quantum gases: new paradigms in many-body
  physics},\ }\href {https://doi.org/10.1080/00018732.2021.1969727} {\bibfield
  {journal} {\bibinfo  {journal} {Advances in Physics}\ }\textbf {\bibinfo
  {volume} {70}},\ \bibinfo {pages} {1} (\bibinfo {year} {2021})}\BibitemShut
  {NoStop}%
\bibitem [{\citenamefont {Baumann}\ \emph {et~al.}(2010)\citenamefont
  {Baumann}, \citenamefont {Guerlin}, \citenamefont {Brennecke},\ and\
  \citenamefont {Esslinger}}]{Baumann2010}%
  \BibitemOpen
  \bibfield  {author} {\bibinfo {author} {\bibfnamefont {K.}~\bibnamefont
  {Baumann}}, \bibinfo {author} {\bibfnamefont {C.}~\bibnamefont {Guerlin}},
  \bibinfo {author} {\bibfnamefont {F.}~\bibnamefont {Brennecke}},\ and\
  \bibinfo {author} {\bibfnamefont {T.}~\bibnamefont {Esslinger}},\ }\bibfield
  {title} {\bibinfo {title} {Dicke quantum phase transition with a superfluid
  gas in an optical cavity},\ }\href {https://doi.org/10.1038/nature09009}
  {\bibfield  {journal} {\bibinfo  {journal} {Nature}\ }\textbf {\bibinfo
  {volume} {464}},\ \bibinfo {pages} {1301} (\bibinfo {year}
  {2010})}\BibitemShut {NoStop}%
\bibitem [{\citenamefont {Klinder}\ \emph
  {et~al.}(2015{\natexlab{a}})\citenamefont {Klinder}, \citenamefont
  {Ke{\ss}ler}, \citenamefont {Wolke}, \citenamefont {Mathey},\ and\
  \citenamefont {Hemmerich}}]{Klinder2015_2}%
  \BibitemOpen
  \bibfield  {author} {\bibinfo {author} {\bibfnamefont {J.}~\bibnamefont
  {Klinder}}, \bibinfo {author} {\bibfnamefont {H.}~\bibnamefont {Ke{\ss}ler}},
  \bibinfo {author} {\bibfnamefont {M.}~\bibnamefont {Wolke}}, \bibinfo
  {author} {\bibfnamefont {L.}~\bibnamefont {Mathey}},\ and\ \bibinfo {author}
  {\bibfnamefont {A.}~\bibnamefont {Hemmerich}},\ }\bibfield  {title} {\bibinfo
  {title} {Dynamical phase transition in the open dicke model},\ }\href
  {https://doi.org/10.1073/pnas.1417132112} {\bibfield  {journal} {\bibinfo
  {journal} {Proceedings of the National Academy of Sciences}\ }\textbf
  {\bibinfo {volume} {112}},\ \bibinfo {pages} {3290} (\bibinfo {year}
  {2015}{\natexlab{a}})}\BibitemShut {NoStop}%
\bibitem [{\citenamefont {Vaidya}\ \emph {et~al.}(2018)\citenamefont {Vaidya},
  \citenamefont {Guo}, \citenamefont {Kroeze}, \citenamefont {Ballantine},
  \citenamefont {Koll\'ar}, \citenamefont {Keeling},\ and\ \citenamefont
  {Lev}}]{Vaidya2018}%
  \BibitemOpen
  \bibfield  {author} {\bibinfo {author} {\bibfnamefont {V.~D.}\ \bibnamefont
  {Vaidya}}, \bibinfo {author} {\bibfnamefont {Y.}~\bibnamefont {Guo}},
  \bibinfo {author} {\bibfnamefont {R.~M.}\ \bibnamefont {Kroeze}}, \bibinfo
  {author} {\bibfnamefont {K.~E.}\ \bibnamefont {Ballantine}}, \bibinfo
  {author} {\bibfnamefont {A.~J.}\ \bibnamefont {Koll\'ar}}, \bibinfo {author}
  {\bibfnamefont {J.}~\bibnamefont {Keeling}},\ and\ \bibinfo {author}
  {\bibfnamefont {B.~L.}\ \bibnamefont {Lev}},\ }\bibfield  {title} {\bibinfo
  {title} {Tunable-range, photon-mediated atomic interactions in multimode
  cavity qed},\ }\href {https://doi.org/10.1103/PhysRevX.8.011002} {\bibfield
  {journal} {\bibinfo  {journal} {Phys. Rev. X}\ }\textbf {\bibinfo {volume}
  {8}},\ \bibinfo {pages} {011002} (\bibinfo {year} {2018})}\BibitemShut
  {NoStop}%
\bibitem [{\citenamefont {Kongkhambut}\ \emph {et~al.}(2022)\citenamefont
  {Kongkhambut}, \citenamefont {Skulte}, \citenamefont {Mathey}, \citenamefont
  {Cosme}, \citenamefont {Hemmerich},\ and\ \citenamefont
  {Ke{\ss}ler}}]{Kongkhambut2022}%
  \BibitemOpen
  \bibfield  {author} {\bibinfo {author} {\bibfnamefont {P.}~\bibnamefont
  {Kongkhambut}}, \bibinfo {author} {\bibfnamefont {J.}~\bibnamefont {Skulte}},
  \bibinfo {author} {\bibfnamefont {L.}~\bibnamefont {Mathey}}, \bibinfo
  {author} {\bibfnamefont {J.~G.}\ \bibnamefont {Cosme}}, \bibinfo {author}
  {\bibfnamefont {A.}~\bibnamefont {Hemmerich}},\ and\ \bibinfo {author}
  {\bibfnamefont {H.}~\bibnamefont {Ke{\ss}ler}},\ }\bibfield  {title}
  {\bibinfo {title} {Observation of a continuous time crystal},\ }\href
  {https://doi.org/10.1126/science.abo3382} {\bibfield  {journal} {\bibinfo
  {journal} {Science}\ }\textbf {\bibinfo {volume} {377}},\ \bibinfo {pages}
  {670} (\bibinfo {year} {2022})}\BibitemShut {NoStop}%
\bibitem [{\citenamefont {Dreon}\ \emph {et~al.}(2022)\citenamefont {Dreon},
  \citenamefont {Baumg{\"a}rtner}, \citenamefont {Hertlein}, \citenamefont
  {Esslinger},\ and\ \citenamefont {Donner}}]{Dreon2022}%
  \BibitemOpen
  \bibfield  {author} {\bibinfo {author} {\bibfnamefont {D.}~\bibnamefont
  {Dreon}}, \bibinfo {author} {\bibfnamefont {X.}~\bibnamefont
  {Baumg{\"a}rtner}, \bibfnamefont {A.and~Li}}, \bibinfo {author}
  {\bibfnamefont {S.}~\bibnamefont {Hertlein}}, \bibinfo {author}
  {\bibfnamefont {T.}~\bibnamefont {Esslinger}},\ and\ \bibinfo {author}
  {\bibfnamefont {T.}~\bibnamefont {Donner}},\ }\bibfield  {title} {\bibinfo
  {title} {Self-oscillating pump in a topological dissipative atom-cavity
  system},\ }\href {https://doi.org/10.1038/s41586-022-04970-0} {\bibfield
  {journal} {\bibinfo  {journal} {Nature}\ }\textbf {\bibinfo {volume} {608}},\
  \bibinfo {pages} {494} (\bibinfo {year} {2022})}\BibitemShut {NoStop}%
\bibitem [{\citenamefont {Kollath}\ \emph {et~al.}(2016)\citenamefont
  {Kollath}, \citenamefont {Sheikhan}, \citenamefont {Wolff},\ and\
  \citenamefont {Brennecke}}]{Kollath2016}%
  \BibitemOpen
  \bibfield  {author} {\bibinfo {author} {\bibfnamefont {C.}~\bibnamefont
  {Kollath}}, \bibinfo {author} {\bibfnamefont {A.}~\bibnamefont {Sheikhan}},
  \bibinfo {author} {\bibfnamefont {S.}~\bibnamefont {Wolff}},\ and\ \bibinfo
  {author} {\bibfnamefont {F.}~\bibnamefont {Brennecke}},\ }\bibfield  {title}
  {\bibinfo {title} {Ultracold fermions in a cavity-induced artificial magnetic
  field},\ }\href {https://doi.org/10.1103/PhysRevLett.116.060401} {\bibfield
  {journal} {\bibinfo  {journal} {Phys. Rev. Lett.}\ }\textbf {\bibinfo
  {volume} {116}},\ \bibinfo {pages} {060401} (\bibinfo {year}
  {2016})}\BibitemShut {NoStop}%
\bibitem [{\citenamefont {Mivehvar}\ \emph {et~al.}(2017)\citenamefont
  {Mivehvar}, \citenamefont {Piazza},\ and\ \citenamefont
  {Ritsch}}]{Mivehvar2017}%
  \BibitemOpen
  \bibfield  {author} {\bibinfo {author} {\bibfnamefont {F.}~\bibnamefont
  {Mivehvar}}, \bibinfo {author} {\bibfnamefont {F.}~\bibnamefont {Piazza}},\
  and\ \bibinfo {author} {\bibfnamefont {H.}~\bibnamefont {Ritsch}},\
  }\bibfield  {title} {\bibinfo {title} {Disorder-driven density and spin
  self-ordering of a bose-einstein condensate in a cavity},\ }\href
  {https://doi.org/10.1103/PhysRevLett.119.063602} {\bibfield  {journal}
  {\bibinfo  {journal} {Phys. Rev. Lett.}\ }\textbf {\bibinfo {volume} {119}},\
  \bibinfo {pages} {063602} (\bibinfo {year} {2017})}\BibitemShut {NoStop}%
\bibitem [{\citenamefont {Bentsen}\ \emph {et~al.}(2019)\citenamefont
  {Bentsen}, \citenamefont {Potirniche}, \citenamefont {Bulchandani},
  \citenamefont {Scaffidi}, \citenamefont {Cao}, \citenamefont {Qi},
  \citenamefont {Schleier-Smith},\ and\ \citenamefont {Altman}}]{Bentsen2019}%
  \BibitemOpen
  \bibfield  {author} {\bibinfo {author} {\bibfnamefont {G.}~\bibnamefont
  {Bentsen}}, \bibinfo {author} {\bibfnamefont {I.-D.}\ \bibnamefont
  {Potirniche}}, \bibinfo {author} {\bibfnamefont {V.~B.}\ \bibnamefont
  {Bulchandani}}, \bibinfo {author} {\bibfnamefont {T.}~\bibnamefont
  {Scaffidi}}, \bibinfo {author} {\bibfnamefont {X.}~\bibnamefont {Cao}},
  \bibinfo {author} {\bibfnamefont {X.-L.}\ \bibnamefont {Qi}}, \bibinfo
  {author} {\bibfnamefont {M.}~\bibnamefont {Schleier-Smith}},\ and\ \bibinfo
  {author} {\bibfnamefont {E.}~\bibnamefont {Altman}},\ }\bibfield  {title}
  {\bibinfo {title} {Integrable and chaotic dynamics of spins coupled to an
  optical cavity},\ }\href {https://doi.org/10.1103/PhysRevX.9.041011}
  {\bibfield  {journal} {\bibinfo  {journal} {Phys. Rev. X}\ }\textbf {\bibinfo
  {volume} {9}},\ \bibinfo {pages} {041011} (\bibinfo {year}
  {2019})}\BibitemShut {NoStop}%
\bibitem [{\citenamefont {J\"ager}\ \emph {et~al.}(2020)\citenamefont
  {J\"ager}, \citenamefont {Holland},\ and\ \citenamefont
  {Morigi}}]{Holland2020}%
  \BibitemOpen
  \bibfield  {author} {\bibinfo {author} {\bibfnamefont {S.~B.}\ \bibnamefont
  {J\"ager}}, \bibinfo {author} {\bibfnamefont {M.~J.}\ \bibnamefont
  {Holland}},\ and\ \bibinfo {author} {\bibfnamefont {G.}~\bibnamefont
  {Morigi}},\ }\bibfield  {title} {\bibinfo {title} {Superradiant
  optomechanical phases of cold atomic gases in optical resonators},\ }\href
  {https://doi.org/10.1103/PhysRevA.101.023616} {\bibfield  {journal} {\bibinfo
   {journal} {Phys. Rev. A}\ }\textbf {\bibinfo {volume} {101}},\ \bibinfo
  {pages} {023616} (\bibinfo {year} {2020})}\BibitemShut {NoStop}%
\bibitem [{\citenamefont {Skulte}\ \emph {et~al.}(2021)\citenamefont {Skulte},
  \citenamefont {Kongkhambut}, \citenamefont {Ke\ss{}ler}, \citenamefont
  {Hemmerich}, \citenamefont {Mathey},\ and\ \citenamefont
  {Cosme}}]{Skulte2021}%
  \BibitemOpen
  \bibfield  {author} {\bibinfo {author} {\bibfnamefont {J.}~\bibnamefont
  {Skulte}}, \bibinfo {author} {\bibfnamefont {P.}~\bibnamefont {Kongkhambut}},
  \bibinfo {author} {\bibfnamefont {H.}~\bibnamefont {Ke\ss{}ler}}, \bibinfo
  {author} {\bibfnamefont {A.}~\bibnamefont {Hemmerich}}, \bibinfo {author}
  {\bibfnamefont {L.}~\bibnamefont {Mathey}},\ and\ \bibinfo {author}
  {\bibfnamefont {J.~G.}\ \bibnamefont {Cosme}},\ }\bibfield  {title} {\bibinfo
  {title} {Parametrically driven dissipative three-level dicke model},\ }\href
  {https://doi.org/10.1103/PhysRevA.104.063705} {\bibfield  {journal} {\bibinfo
   {journal} {Phys. Rev. A}\ }\textbf {\bibinfo {volume} {104}},\ \bibinfo
  {pages} {063705} (\bibinfo {year} {2021})}\BibitemShut {NoStop}%
\bibitem [{\citenamefont {Rosa-Medina}\ \emph {et~al.}(2022)\citenamefont
  {Rosa-Medina}, \citenamefont {Ferri}, \citenamefont {Finger}, \citenamefont
  {Dogra}, \citenamefont {Kroeger}, \citenamefont {Lin}, \citenamefont
  {Chitra}, \citenamefont {Donner},\ and\ \citenamefont
  {Esslinger}}]{Rodrigo2022}%
  \BibitemOpen
  \bibfield  {author} {\bibinfo {author} {\bibfnamefont {R.}~\bibnamefont
  {Rosa-Medina}}, \bibinfo {author} {\bibfnamefont {F.}~\bibnamefont {Ferri}},
  \bibinfo {author} {\bibfnamefont {F.}~\bibnamefont {Finger}}, \bibinfo
  {author} {\bibfnamefont {N.}~\bibnamefont {Dogra}}, \bibinfo {author}
  {\bibfnamefont {K.}~\bibnamefont {Kroeger}}, \bibinfo {author} {\bibfnamefont
  {R.}~\bibnamefont {Lin}}, \bibinfo {author} {\bibfnamefont {R.}~\bibnamefont
  {Chitra}}, \bibinfo {author} {\bibfnamefont {T.}~\bibnamefont {Donner}},\
  and\ \bibinfo {author} {\bibfnamefont {T.}~\bibnamefont {Esslinger}},\
  }\bibfield  {title} {\bibinfo {title} {Observing dynamical currents in a
  non-hermitian momentum lattice},\ }\href
  {https://doi.org/10.1103/PhysRevLett.128.143602} {\bibfield  {journal}
  {\bibinfo  {journal} {Phys. Rev. Lett.}\ }\textbf {\bibinfo {volume} {128}},\
  \bibinfo {pages} {143602} (\bibinfo {year} {2022})}\BibitemShut {NoStop}%
\bibitem [{\citenamefont {Zhang}\ \emph {et~al.}(2022)\citenamefont {Zhang},
  \citenamefont {Dreon}, \citenamefont {Esslinger}, \citenamefont {Jaksch},
  \citenamefont {Buca},\ and\ \citenamefont {Donner}}]{Zhang2022}%
  \BibitemOpen
  \bibfield  {author} {\bibinfo {author} {\bibfnamefont {Z.}~\bibnamefont
  {Zhang}}, \bibinfo {author} {\bibfnamefont {D.}~\bibnamefont {Dreon}},
  \bibinfo {author} {\bibfnamefont {T.}~\bibnamefont {Esslinger}}, \bibinfo
  {author} {\bibfnamefont {D.}~\bibnamefont {Jaksch}}, \bibinfo {author}
  {\bibfnamefont {B.}~\bibnamefont {Buca}},\ and\ \bibinfo {author}
  {\bibfnamefont {T.}~\bibnamefont {Donner}},\ }\href
  {https://doi.org/10.48550/ARXIV.2205.01461} {\bibinfo {title} {Tunable
  non-equilibrium phase transitions between spatial and temporal order through
  dissipation}} (\bibinfo {year} {2022})\BibitemShut {NoStop}%
\bibitem [{\citenamefont {De~Bernardis}\ \emph {et~al.}(2021)\citenamefont
  {De~Bernardis}, \citenamefont {Cian}, \citenamefont {Carusotto},
  \citenamefont {Hafezi},\ and\ \citenamefont {Rabl}}]{Bernardis2021}%
  \BibitemOpen
  \bibfield  {author} {\bibinfo {author} {\bibfnamefont {D.}~\bibnamefont
  {De~Bernardis}}, \bibinfo {author} {\bibfnamefont {Z.-P.}\ \bibnamefont
  {Cian}}, \bibinfo {author} {\bibfnamefont {I.}~\bibnamefont {Carusotto}},
  \bibinfo {author} {\bibfnamefont {M.}~\bibnamefont {Hafezi}},\ and\ \bibinfo
  {author} {\bibfnamefont {P.}~\bibnamefont {Rabl}},\ }\bibfield  {title}
  {\bibinfo {title} {Light-matter interactions in synthetic magnetic fields:
  Landau-photon polaritons},\ }\href
  {https://doi.org/10.1103/PhysRevLett.126.103603} {\bibfield  {journal}
  {\bibinfo  {journal} {Phys. Rev. Lett.}\ }\textbf {\bibinfo {volume} {126}},\
  \bibinfo {pages} {103603} (\bibinfo {year} {2021})}\BibitemShut {NoStop}%
\bibitem [{\citenamefont {Gauthier}\ \emph {et~al.}(2016)\citenamefont
  {Gauthier}, \citenamefont {Lenton}, \citenamefont {McKay~Parry},
  \citenamefont {Baker}, \citenamefont {Davis}, \citenamefont
  {Rubinsztein-Dunlop},\ and\ \citenamefont {Neely}}]{Gauthier2016}%
  \BibitemOpen
  \bibfield  {author} {\bibinfo {author} {\bibfnamefont {G.}~\bibnamefont
  {Gauthier}}, \bibinfo {author} {\bibfnamefont {I.}~\bibnamefont {Lenton}},
  \bibinfo {author} {\bibfnamefont {N.}~\bibnamefont {McKay~Parry}}, \bibinfo
  {author} {\bibfnamefont {M.}~\bibnamefont {Baker}}, \bibinfo {author}
  {\bibfnamefont {M.~J.}\ \bibnamefont {Davis}}, \bibinfo {author}
  {\bibfnamefont {H.}~\bibnamefont {Rubinsztein-Dunlop}},\ and\ \bibinfo
  {author} {\bibfnamefont {T.~W.}\ \bibnamefont {Neely}},\ }\bibfield  {title}
  {\bibinfo {title} {Direct imaging of a digital-micromirror device for
  configurable microscopic optical potentials},\ }\href
  {https://doi.org/10.1364/optica.3.001136} {\bibfield  {journal} {\bibinfo
  {journal} {Optica}\ }\textbf {\bibinfo {volume} {3}},\ \bibinfo {pages}
  {1136} (\bibinfo {year} {2016})}\BibitemShut {NoStop}%
\bibitem [{\citenamefont {Bloch}\ \emph {et~al.}(2008)\citenamefont {Bloch},
  \citenamefont {Dalibard},\ and\ \citenamefont {Zwerger}}]{Bloch2008}%
  \BibitemOpen
  \bibfield  {author} {\bibinfo {author} {\bibfnamefont {I.}~\bibnamefont
  {Bloch}}, \bibinfo {author} {\bibfnamefont {J.}~\bibnamefont {Dalibard}},\
  and\ \bibinfo {author} {\bibfnamefont {W.}~\bibnamefont {Zwerger}},\
  }\bibfield  {title} {\bibinfo {title} {Many-body physics with ultracold
  gases},\ }\href {https://doi.org/10.1103/RevModPhys.80.885} {\bibfield
  {journal} {\bibinfo  {journal} {Rev. Mod. Phys.}\ }\textbf {\bibinfo {volume}
  {80}},\ \bibinfo {pages} {885} (\bibinfo {year} {2008})}\BibitemShut
  {NoStop}%
\bibitem [{\citenamefont {Cosme}\ \emph {et~al.}(2019)\citenamefont {Cosme},
  \citenamefont {Skulte},\ and\ \citenamefont {Mathey}}]{Cosme2019}%
  \BibitemOpen
  \bibfield  {author} {\bibinfo {author} {\bibfnamefont {J.~G.}\ \bibnamefont
  {Cosme}}, \bibinfo {author} {\bibfnamefont {J.}~\bibnamefont {Skulte}},\ and\
  \bibinfo {author} {\bibfnamefont {L.}~\bibnamefont {Mathey}},\ }\bibfield
  {title} {\bibinfo {title} {Time crystals in a shaken atom-cavity system},\
  }\href {https://doi.org/10.1103/PhysRevA.100.053615} {\bibfield  {journal}
  {\bibinfo  {journal} {Phys. Rev. A}\ }\textbf {\bibinfo {volume} {100}},\
  \bibinfo {pages} {053615} (\bibinfo {year} {2019})}\BibitemShut {NoStop}%
\bibitem [{\citenamefont {Skulte}\ \emph {et~al.}(2022)\citenamefont {Skulte},
  \citenamefont {Kongkhambut}, \citenamefont {Rao}, \citenamefont {Keßler},
  \citenamefont {Hemmerich},\ and\ \citenamefont {Cosme}}]{Skulte2022}%
  \BibitemOpen
  \bibfield  {author} {\bibinfo {author} {\bibfnamefont {J.}~\bibnamefont
  {Skulte}}, \bibinfo {author} {\bibfnamefont {P.}~\bibnamefont {Kongkhambut}},
  \bibinfo {author} {\bibfnamefont {L.}~\bibnamefont {Rao}, \bibfnamefont
  {S.and~Mathey}}, \bibinfo {author} {\bibfnamefont {H.}~\bibnamefont
  {Keßler}}, \bibinfo {author} {\bibfnamefont {A.}~\bibnamefont {Hemmerich}},\
  and\ \bibinfo {author} {\bibfnamefont {J.~G.}\ \bibnamefont {Cosme}},\ }\href
  {https://doi.org/10.48550/ARXIV.2209.03342} {\bibinfo {title} {Condensate
  formation in a dark state of a driven atom-cavity system}} (\bibinfo {year}
  {2022})\BibitemShut {NoStop}%
\bibitem [{\citenamefont {Klinder}\ \emph
  {et~al.}(2015{\natexlab{b}})\citenamefont {Klinder}, \citenamefont
  {Ke\ss{}ler}, \citenamefont {Bakhtiari}, \citenamefont {Thorwart},\ and\
  \citenamefont {Hemmerich}}]{Klinder2015}%
  \BibitemOpen
  \bibfield  {author} {\bibinfo {author} {\bibfnamefont {J.}~\bibnamefont
  {Klinder}}, \bibinfo {author} {\bibfnamefont {H.}~\bibnamefont {Ke\ss{}ler}},
  \bibinfo {author} {\bibfnamefont {M.~R.}\ \bibnamefont {Bakhtiari}}, \bibinfo
  {author} {\bibfnamefont {M.}~\bibnamefont {Thorwart}},\ and\ \bibinfo
  {author} {\bibfnamefont {A.}~\bibnamefont {Hemmerich}},\ }\bibfield  {title}
  {\bibinfo {title} {Observation of a superradiant mott insulator in the
  dicke-hubbard model},\ }\href
  {https://doi.org/10.1103/PhysRevLett.115.230403} {\bibfield  {journal}
  {\bibinfo  {journal} {Phys. Rev. Lett.}\ }\textbf {\bibinfo {volume} {115}},\
  \bibinfo {pages} {230403} (\bibinfo {year} {2015}{\natexlab{b}})}\BibitemShut
  {NoStop}%
\bibitem [{sup()}]{suppmat}%
  \BibitemOpen
  \href@noop {} {}\bibinfo {note} {{See Supplemental Material for more details
  on the stability against particle number fluctuations.}}\BibitemShut {Stop}%
\bibitem [{\citenamefont {Landig}\ \emph {et~al.}(2016)\citenamefont {Landig},
  \citenamefont {Hruby}, \citenamefont {Dogra}, \citenamefont {Landini},
  \citenamefont {Mottl}, \citenamefont {Donner},\ and\ \citenamefont
  {Esslinger}}]{Landig2016}%
  \BibitemOpen
  \bibfield  {author} {\bibinfo {author} {\bibfnamefont {R.}~\bibnamefont
  {Landig}}, \bibinfo {author} {\bibfnamefont {L.}~\bibnamefont {Hruby}},
  \bibinfo {author} {\bibfnamefont {N.}~\bibnamefont {Dogra}}, \bibinfo
  {author} {\bibfnamefont {M.}~\bibnamefont {Landini}}, \bibinfo {author}
  {\bibfnamefont {R.}~\bibnamefont {Mottl}}, \bibinfo {author} {\bibfnamefont
  {T.}~\bibnamefont {Donner}},\ and\ \bibinfo {author} {\bibfnamefont
  {T.}~\bibnamefont {Esslinger}},\ }\bibfield  {title} {\bibinfo {title}
  {Quantum phases from competing short- and long-range interactions in an
  optical lattice},\ }\href {https://doi.org/10.1038/nature17409} {\bibfield
  {journal} {\bibinfo  {journal} {Nature}\ }\textbf {\bibinfo {volume} {532}},\
  \bibinfo {pages} {476} (\bibinfo {year} {2016})}\BibitemShut {NoStop}%
\end{thebibliography}%


\providecommand{\noopsort}[1]{}\providecommand{\singleletter}[1]{#1}%
\begin{thebibliography}{0}%
\makeatletter
\providecommand \@ifxundefined [1]{%
 \@ifx{#1\undefined}
}%
\providecommand \@ifnum [1]{%
 \ifnum #1\expandafter \@firstoftwo
 \else \expandafter \@secondoftwo
 \fi
}%
\providecommand \@ifx [1]{%
 \ifx #1\expandafter \@firstoftwo
 \else \expandafter \@secondoftwo
 \fi
}%
\providecommand \natexlab [1]{#1}%
\providecommand \enquote  [1]{``#1''}%
\providecommand \bibnamefont  [1]{#1}%
\providecommand \bibfnamefont [1]{#1}%
\providecommand \citenamefont [1]{#1}%
\providecommand \href@noop [0]{\@secondoftwo}%
\providecommand \href [0]{\begingroup \@sanitize@url \@href}%
\providecommand \@href[1]{\@@startlink{#1}\@@href}%
\providecommand \@@href[1]{\endgroup#1\@@endlink}%
\providecommand \@sanitize@url [0]{\catcode `\\12\catcode `\$12\catcode
  `\&12\catcode `\#12\catcode `\^12\catcode `\_12\catcode `\%12\relax}%
\providecommand \@@startlink[1]{}%
\providecommand \@@endlink[0]{}%
\providecommand \url  [0]{\begingroup\@sanitize@url \@url }%
\providecommand \@url [1]{\endgroup\@href {#1}{\urlprefix }}%
\providecommand \urlprefix  [0]{URL }%
\providecommand \Eprint [0]{\href }%
\providecommand \doibase [0]{https://doi.org/}%
\providecommand \selectlanguage [0]{\@gobble}%
\providecommand \bibinfo  [0]{\@secondoftwo}%
\providecommand \bibfield  [0]{\@secondoftwo}%
\providecommand \translation [1]{[#1]}%
\providecommand \BibitemOpen [0]{}%
\providecommand \bibitemStop [0]{}%
\providecommand \bibitemNoStop [0]{.\EOS\space}%
\providecommand \EOS [0]{\spacefactor3000\relax}%
\providecommand \BibitemShut  [1]{\csname bibitem#1\endcsname}%
\let\auto@bib@innerbib\@empty
\end{thebibliography}%

\end{document}